\documentclass[twocolumn,longauthor,usenames,dvipsnames]{aastex63}

\newbox\grsign \setbox\grsign=\hbox{$>$}
\newdimen\grdimen \grdimen=\ht\grsign
\newbox\laxbox \newbox\gaxbox
\setbox\gaxbox=\hbox{\raise.5ex\hbox{$>$}\llap
     {\lower.5ex\hbox{$\sim$}}}\ht1=\grdimen\dp1=0pt
\setbox\laxbox=\hbox{\raise.5ex\hbox{$<$}\llap
     {\lower.5ex\hbox{$\sim$}}}\ht2=\grdimen\dp2=0pt

\usepackage[T1]{fontenc}
\usepackage{amsmath}
\usepackage{amssymb}
\usepackage{ctable} 
\usepackage{graphicx}
\usepackage{natbib}
\usepackage[titletoc,title]{appendix}
\usepackage{ulem}

\usepackage{mathtools}
\usepackage{lmodern}
\usepackage{csquotes}
\usepackage{xspace}
 
\begin{document}

\title{Curved jet motion. I. Orbiting and precessing jets}
\shorttitle{Curved jets}
\shortauthors{Fendt \& Yardimci}

\author[0000-0002-3528-7625]{Christian Fendt}
\affiliation{Max Planck Institute for Astronomy, Heidelberg, Germany; fendt@mpia.de}

\author[0000-0001-8697-2385]{Melis Yardimci}
\affiliation{University of Ege, Department of Astronomy and Space Sciences, 35100, Izmir, Turkey}
\affiliation{Max Planck Institute for Astronomy, Heidelberg, Germany; yardimci@mpia.de}

\begin{abstract}
Astrophysical jets are often observed as bent or curved structures.
We also know that the different jet sources may be binary in nature, which may lead to a regular, periodic motion of the
jet nozzle, an orbital motion or precession.
Here, we present the results of 2D (M)HD simulations in order to investigate how a precessing or orbiting jet nozzle
affects the propagation of a high-speed jet.
We have performed a parameter study of systems with different precession angles, orbital periods or separations, and different 
magnetic field strengths.
We find that these kinds of nozzles lead to curved jet propagation which is determined by the main parameters that define 
the jet nozzle. 
We find C-shaped jets from orbiting nozzles and S-shaped jets from precessing nozzles.
Over long time and long distances, the initially curved jet motion bores a broad channel into the ambient gas that is filled 
with high-speed jet material which lateral motion is damped, however.
A strong (longitudinal) magnetic field can damp the jet curvature that is enforced by either precession of orbital
motion of the jet sources.
We have investigated the force balance across the jet and ambient medium and found that the lateral magnetic pressure and gas pressure 
gradients are almost balanced, but that a lack of gas pressure on the concave side of the curvature is leading to the lateral
motion. 
Magnetic tension does not play a significant role.
Our results are obtained in code units, but we provide scaling relations such that our results may be applied to young stars, micro-quasars,
symbiotic stars or AGN.
\end{abstract}


\section{Introduction}
\label{sec:intro}
Astrophysical jets as highly collimated, high-speed outflows of material are launched from a wide range of 
astrophysical objects.
Jets are observed from accreting brown dwarfs, young stars (YSO), symbiotic stars (SySt), microquasars and 
active galactic nuclei (AGN) spanning orders of magnitude in energy output and length scale.
It is accepted today that jets are produced by an interplay of a
large-scale magnetic field with an accretion disk 
\citep{1982MNRAS.199..883B, 1983ApJ...274..677P, 1985PASJ...37..515U, 1997A&A...319..340F,
2002ApJ...581..988C,2007A&A...469..811Z, 2015SSRv..191..441H,2016ApJ...825...14S}.

There are well-known observational signatures that strongly indicate on non-axisymmetric features like jet 
precession or a curved ballistic motion of the jet which are suggesting that the jet source is part of a binary 
system or even a multiple system
\citep{Fendt1998, 2000ApJ...535..833S, 2002MNRAS.335.1100C, 2004HEAD....8.2903M, 2007A&A...476L..17A, 2014xru..confE.147M,
2016A&A...593A.132P, Beltran2016, 2019ASSP...55...71M,2019A&A...622L...3E, 2019IAUS..346...34M, 2021MNRAS.503..704M, 2021MNRAS.503.3145B, 2021MNRAS.tmp..799D}.

In \citet{Fendt1998} a number of possible mechanisms that can bend the jet away from its straight motion were 
investigated phenomenologically, in particular electromagnetic forces, an orbital motion, or side winds (by a companion 
stellar wind, or the proper motion of the jet source).

The propagation of precessing jets has been modeled in some detail already early on.
Analytical studies of jets with time-dependent direction of injection \citep{1993MNRAS.260..163R}
have revealed the principal large-scale structure of these flows, namely the {"}garden hose{"} effect,
a helical structure of the jet with a wavelength decreasing with distance and an amplitude
that is increasing.
Precessing jets were further studied in 2D slab jet simulations that were confirming this picture \citep{1995MNRAS.275..557B}.
Due to the time-dependent simulation setup, also features such as the bow shock structure or the interaction with the
ambient gas could be studied.

Studies in 3D are often thought to be more realistic, however, they are computationally very expensive. 
Due to the low numerical resolution that is sometimes applied, the reality of some of the studies may be disputed.
The first studies of this kind were presented in seminal studies by \citet{1996MNRAS.282.1114C}, clearly indicating a 
3D helical jet structure, with more or less helical windings evolving, depending on the model parameters for jet injection.
With the resolution applied features like shock layers or internal jet structures could not yet be resolved.
The simulations of supersonic by \citet{1996MNRAS.282.1114C} were performed on a Cartesian grid (of $128^3$ or$256^3$ 
grid cells respectively), applying jet precession angles of about 20 degrees.

Simulations in 3D hydrodynamics of orbiting jets were presented by \citet{2002ApJ...568..733M} in principle confirming 
the predictions of \citet{Fendt1998} of S-shaped bipolar jets from orbiting jet sources.
The simulations in addition showed a widening of the overall jet channel, similar to the analytical precession
model of \citet{1993MNRAS.260..163R}.
These features can be relevant and have been applied, for hypothetical jet outflows launched by circum-{\em planetary} 
accretion disks in young planet-forming stars \citep{2003A&A...411..623F}.
 
Relativistic 3D hydrodynamic simulations of the SS433 jet propagation have been performed 
by \citet{2015A&A...574A.143M}, indicating a smooth transition
from a precessing jet nozzle into a hollow non-precessing jet on larger scales.

To our knowledge, since then no simulation study was published that was investigating quantitatively and 
applying high resolution the impact of precession or orbital motion on the jet propagation.
Our present paper aims to investigate these effects, in particular also the impact of magnetic fields.

Here we constrain ourselves to 2D simulations that allow to investigate a series of parameter runs 
applying different injection models, and to investigate how they affect the global structure of the 
jet.
We rely on the grid geometry of a slab jet (as e.g. \citealt{1995MNRAS.275..557B}).
It is essential for this study to break the axisymmetry assumption, since the jet beam changes direction
over the quadrants situated {"}left{"} and {"}right{"} the precession axis  or the axis of orbital
motion.
Therefore jet propagation we obtain in our 2D approach should be thought as representing the 
3D jet propagation {\em projected} onto the plane of the sky.
Further studies applying a full 3D modeling will be published in a future study.

The paper is organized as follows. 
In Sect.~\ref{sec:obs}  we provide a brief overview of curved jet motions for a variety of sources.
In Sect.~\ref{sec:model} we describe our model setup. 
We present our simulation results on small scales in Sect.~\ref{sec:nozzles} discussing
various model setups.
This is then extended to intermediate time scales and length scales in Sect.~\ref{sec:c-shape} and Sect.~\ref{sec:s-shape}.
In Sect.~\ref{sec:channel} we discuss the jet evolution for very long time scales.
Sect.~\ref{sec:magnfield} is  considering the influence of the magnetic field on the jet
propagation.
In Sect.~\ref{sec:sum} we summarize our work.
We provide a comparison of our approach with selected literature work in Appendix~\ref{appendix:comparison}.

In a follow-up paper, we will present simulations of jet motion under the influence of a side wind
and discuss them in the astrophysical context of symbiotic stars.

\section{Observations of curved jets}
\label{sec:obs}
Curved or asymmetric jet motion typically arises when jets are launched in binaries.
Other causes may be side winds by the ambient medium or by a companion star.
Observational evidence for the features just mentioned has been found for a number of sources.

We now know that stars typically form as binaries or higher order multiple systems
\citep{2007ARA&A..45..565M}.
Among the binary systems that launch jets, are T\,Tau
\citep{1997A&AS..126..437H, 2002ApJ...568..771D, 2003AJ....125..858J}
or RW\,Aur
\citep{1996AJ....111.2403H, 2012ARep...56..686B}.
Non-axisymmetric jet motion is observed from the binary system HK\,Tau \citep{2014Natur.511..567J},
suggesting that one or both of the stellar disks may be inclined to the orbital
plane \citep{2000MNRAS.317..773B}.
Another exiting example is such as L1157 \citep{2016A&A...593L...4P} that shows a clear non-axisymmetric, 
S-shaped propagation pattern.

The pre-main sequence binary KH\,15D \citet{2010ApJ...708L...5M} shows a spectroscopically identified bipolar jet.
This seems to be launched from the innermost part of the circum-binary disk, or may, alternatively, result from 
merging two outflows driven by the individual stars, respectively.
This system is now known to be a young ($\sim 3$ Myr) eccentric binary system embedded in a nearly
edge-on circum-binary disk that is tilted, warped, and is precessing with respect to the binary orbit
\citep{2004AJ....127.2344J, 2004ApJ...607..913C, 2004AJ....128.1265J, 2004ApJ...603L..45W}.

The precessing jet of SS 433 is probably the best understood astrophysical precessing jet \citep{1979ApJ...233L..63M, 2015A&A...574A.143M}. 
This system is a microquasar, sharing some similarities with symbiotic stars, e.g. CH\,Cyg and MWC\,560 \citep{Zamanov2005}.   

Another prominent class, although diverse, is the class of symbiotic stars. 
Here, as well, if a jet is ejected, the intrinsic binarity of the source will naturally lead to asymmetric 
effects of jet propagation.
Symbiotic stars are systems closely related to cataclysmic variables. 
Both are binary systems sharing some common features, such as having the same type of compact star, a white dwarf, 
and also show similarities on the X-ray spectrum and/or in the outburst behavior in the optical regime. 
In spite of these similarities\footnote{However, note that their orbital separations are inversely related.}, 
there is little evidence for cataclysmic variables having jets. 

In comparison, for the cataclysmic variable U~Sco of the subclass of recurrent nova 
the observed outflow velocity reaches $\simeq 5000\,{\rm km s^{-1}}$ \citep{Kato03}.

A typical example for a symbiotic binary system that shows a bipolar outflow is R Aquarii (R~Aqr).
It is surrounded by an hourglass-like bipolar shell up to scales of $40{''} $ \citep{Solf85}.
The axis of these bipolar shells is inclined at $18\degr$ with respect to the plane of the sky. 
Interestingly, R~Aqr shows a characteristic S-shape structure of the inner outflow in the radio \citep{Hollis93}
as well as in the optical [OIII] $\lambda 5007$ emission \citep{Melni2018}. 

An additional feature, that is rather common for symbiotic stars, but not for the other jet sources discussed 
so far, is the existence of strong, cold stellar winds in these systems -- probably launched from the giant 
stellar component. 
It would be therefore interesting to model how such stellar winds affect the propagation of the jet stream.
Overall, this is another 3D effect that arises due to the binary nature of these sources.
In a follow-up paper, we will investigate the influence of side winds in a 2D approximation.

Binarity of the jet source and subsequent jet precession is also discussed for extra-galactic jets
\citep{1980Natur.287..307B, 2013A&A...557A..85R, 2014MNRAS.437...32W, 2017NatAs...1..727K, 2019ApJ...873...11J}.
A prototype of these kinds of active galactic nuclei is the blazar OJ~287. The periodicity in the light curve
of about 11 years is usually explained by a central binary supermassive black hole 
\citep{1988ApJ...325..628S,2016ApJ...819L..37V}.
Recent modeling concluded that  the observed jet structure and periodicity could be explained by
a precessing jet which launching engine is influenced by tidal forces by the secondary black hole binary \citep{2018MNRAS.478.3199B}.

In the following we will investigate how a variety in the jet launching parameters such as precession angle, binary orbit,
jet internal and external Mach number, or the magnetic field strength affects the curved propagation of the outflow.

\section{Model approach}
\label{sec:model}
We apply the magnetohydrodynamic (MHD) code PLUTO \citep{2007ApJS..170..228M}, version 4.3,
when solving the time-dependent MHD equations on a Cartesian grid, (x,y,z).
We apply the geometry of a two-dimensional {\em slab},
meaning that we neglect field components and derivatives in the third direction,
here chosen as the $z$-direction.

Also, the gravity of the jet source of point mass $M$ is neglected,
as we know from jet launching simulations \citep{2014ApJ...793...31S} that disk winds are accelerated
to super-escape speed very quickly.
The code numerically solves the equations for the mass conservation,
\begin{equation}
\frac{\partial\rho}{\partial t} + \nabla \cdot(\rho \vec{v})=0,
\end{equation}
with the mass density $\rho$ and flow velocity $\vec{v}$,
the momentum conservation,
\begin{equation}
\frac{\partial \rho \vec{v}}{\partial t} + 
\nabla \cdot \left[ \rho  \vec{v} \vec{v} +
\left(P + \frac{\vec{B} \cdot \vec{B}}{2}\right) I - 
\vec{B}\vec{B} \right] = 0
\label{eq:momentum}
\end{equation}
with the thermal pressure $P$ and the magnetic field $\vec{B}$. 
Note that here the equations are written in non-dimensional form, and, as usual, the
factor $4\pi$ is incorporated in the definition of the magnetic field.
We apply a polytropic equation of state,
$P \propto \rho^{\gamma}$,
with the polytropic index $\gamma=5/3$.

The code further solves for the conservation of energy,
\begin{equation}
\frac{\partial e}{\partial t} +  \nabla \cdot \left[ \left( e + P + \frac{\vec{B} \cdot \vec{B}}{2} \right) \vec{v} - \left(\vec{v} \cdot \vec{B} \right) \vec{B} \right] = -\Lambda_{\rm{cool}},
\end{equation}
with the total energy density,
\begin{equation}
e = \frac{P}{\gamma -1} + 
    \frac{\rho \vec{v} \cdot \vec{v}}{2} + 
    \frac{\vec{B} \cdot \vec{B}}{2},
\label{eq:energy}
\end{equation}
given by the sum of thermal, kinetic, magnetic energy, respectively. 
In Equation~\ref{eq:energy} we have neglected resistive heating, and we will further neglect the cooling term on the r.h.s.
The electric current density is denoted by $\vec{j}=\nabla \times \vec{B}$.
The magnetic field evolution is governed by the induction equation
\begin{equation}
\frac{\partial \vec{B}}{\partial t} =  \nabla\times \left( \vec{v} \times \vec{B} \right),
\end{equation}
where we again have neglected the resistive term.

We use a Harten-Lax-van Leer (HLL) Riemann solver together with a third-order order 
Runge-Kutta scheme for time evolution and the PPM (piecewise parabolic method) reconstruction 
of \citep{1984JCoPh..54..174C} for spatial integration.
The magnetic field evolution follows the method of Constrained Transport \citep{2004JCoPh.195...17L}.
We satisfy the Courant-Friedrichs-Levy condition by using a CFL number of 0.4.

\begin{table}
\caption{Astrophysical scales of various jet parameters for different sources, such as YSOs, MQs, SySts, and AGNs.
Shown are  typical values for the jet radius $R_{\rm jet}$, the jet density $\rho_{\rm jet}$,
the jet speed $V_0$, and the magnetic field strength involved $B_0$.
The timescale follows from $T_0 = L_0 / V_0 = R_{\rm jet}/V_{\rm jet}$, where we assume the jet radius as the typical length scale.
Also, typical mass fluxes $\dot{M}_{\rm jet}$ and energy fluxes $\dot{E}_{\rm jet}$ are indicated. }
\begin{center}
    \begin{tabular}{llllll}
    ~                 & YSO  & MQ & SySt     &AGN  & [unit] \\     
    \noalign{\smallskip}    \hline \hline    \noalign{\smallskip}
    $R_{\rm jet}$     & 50   & 1  &  $\simeq$ 1    &1000 & au   \\
    $\rho_{\rm jet}$  & 1000 & 1  &  $10^2-10^4$        &1    & $\rm {cm}^{-3}$  \\    
    \noalign{\smallskip}\hline \noalign{\smallskip}
    $V_0$     & 100   & $\simeq$\,c & 250-1000 & $\simeq$\,c& $\rm km\,s^{-1}$ \\
    $B_0$     & $10^{-4}$         &0.5          &  $10^{-1}$  & 1000       & G       \\
    $T_0$     & 1.7               &0.25         &   0.1-0.4   & 0.5        & days      \\
   $\dot{M}_{\rm jet}$ &       $ 10^{-5}$   &        $10^{-10}$& $10^{-8}-10^{-9}$ & $10$ & $M_{\odot}{\rm yr}^{-1}$\\ 
   $\dot{E}_{\rm jet}$ &          $10^{35}$ &          $10^{30}$ & $10^{33}$   &   $10^{46}$ & ${\rm ergs\,s}^{-1}$   
    \end{tabular}
\end{center}
\label{tbl:normalization}
\end{table}

\begin{table} 
\caption{Numerical setup and parameters for the simulations of precessing and orbiting jets.
Shown is the orbital period $P_{\rm orb}$, the binary separation $a$ and the precession period $P_{\rm prec}$. 
The grid size is denoted by e.g. x40y200, referring to $[-40 < x < 40]$, similar for the $y$-range,
similar for the $x$-range. 
All simulations have the same resolution, $\Delta x = \Delta y = 0.05$. 
For a comparison of our parameters with selected literature work see Appendix~\ref{appendix:comparison}.
} 
\begin{center}
    \begin{tabular}{ccccc} 
    \multicolumn{5}{l}{Orbiting jets:} \\
    \noalign{\smallskip}
    \hline \hline
    \noalign{\smallskip}
   Simulation ID & $\beta$ & $P_{\rm orb}$ & $a$ & Grid size \\
   \noalign{\smallskip}
    \hline
    \noalign{\smallskip}
    {\em ojP10a5}      & HD    & 10        & 5    & x40y200 \\
    {\em ojP25a10}     & HD    & 25        & 10   & x80y200\\
    {\em ojP25a5}      & HD    & 25        & 5    & x40y200\\
    {\em ojP25a10B100} & 100   & 25        & 10   & x80y200 \\
    {\em ojP25a10B10}  & 10    & 25        & 10   & x80y200 \\
    {\em ojP25a10B1}   & 1     & 25        & 10   & x80y200 \\
    \noalign{\smallskip}
    \hline
     & & & &   \\
     \multicolumn{5}{l}{Precessing nozzles:} \\
     \noalign{\smallskip}
    \hline \hline
    \noalign{\smallskip}
    Simulation ID & $\beta$ & $\alpha_{\rm max}$ & $P_{prec}$ & Grid size  \\
    \noalign{\smallskip}
    \hline
    \noalign{\smallskip}
    {\em pj10deg}      & HD      & 10$\degr$ & 10     & x40y200   \\ 
    {\em pj30deg}      & HD      & 30$\degr$ & 10     & x40y200   \\
    {\em pj45deg}      & HD      & 45$\degr$ & 10     & x40y200   \\
    {\em pj60deg}      & HD      & 60$\degr$ & 10     & x40y100   \\
    {\em pj75deg}      & HD      & 75$\degr$ & 10     & x20y100   \\
    {\em pj30degB100}  & 100     & 30$\degr$ & 10     & x30y100   \\
    {\em pj30degB10}   & 10      & 30$\degr$ & 10     & x20y100   \\
    {\em pj30degB1}    & 1       & 30$\degr$ & 10     & x40y200   \\
    \noalign{\smallskip}
    \hline  
    \end{tabular}
\end{center}
\label{tbl:box_sizes}
\end{table}

\subsection{Normalization and numerical grid}
\label{sec:numgrid}
No physical scales are introduced in the equations above ab initio. 
Therefore, we will present the results of our simulations in non-dimensional code units,
that can be re-scaled to different astrophysical sources.

For the simulation setup,
lengths are given in units of the jet radius $R_{\rm jet}$, thus $\hat{x}_{\rm jet} = 1.0$.
The jet is injected with a jet speed $\hat{v}_{\rm jet}$ in code units.
The time unit $t_{\rm jet}$ corresponds to the dynamical timescale of the jet 
$t_{\rm dyn} \equiv \hat{x}_{\rm jet} / \hat{v}_{\rm jet} $. 
Densities are given in code units, $\hat{\rho}_{\rm jet}$ for the jet and $\hat{\rho}_{\rm ism}$ 
for the ambient medium.

Typically, we inject an over-dense jet $\hat{\rho}_{\rm jet} = 1$ into an ambient medium with $\hat{\rho}_{\rm ism} = 0.1$,
corresponding to a density contrast of $\eta \equiv \hat{\rho}_{\rm jet} / \hat{\rho}_{\rm ism} = 10$.
The gas pressures of jet and ambient medium are in equilibrium, $\hat{P}_{\rm jet} = \hat{P}_{\rm ism} = 0.6$.
For the number value $P=0.6$ we follow classical papers \citep{1993ApJ...413..198S, 1993ApJ...413..210S}, 
which provides an external Mach number $M_{\rm ext} = 3.16$, 
respectively an internal Mach number of $M_{\rm int} = 10$, applying a jet speed of $\hat{v}_{\rm jet} = 10$.
In the following we will omit the hats, referring always to code units, if not stated otherwise.

Table~\ref{tbl:normalization} summarizes the typical astrophysical scales for the leading
physical variables for four different object classes -
Young Stellar Objects (YSOs), Symbiotic Stars (SySts), micro-quasars (MQs), and Active Galactic Nuclei (AGNs). 
Simulation scales in code units can be re-scaled to astrophysical units accordingly.

We apply a numerical grid with equidistant spacing in $y$-direction from $y = [0, y_{\rm out}]$.
In $x$-direction we apply a high resolution for the close jet environment $|x| < 10$,
and a scaled grid for $10 < |x| < |x_{\rm out}|$.

Our typical computational domain has a physical size of $-40 < x < 40$, and $0 < y < 200$,
and is discretized with $n_x \times n_y$ grid cells, usually $n_x = 400, n_y = 2000$.
We have also done a resolution study with higher resolution.
In order to avoid interference of the jets with the outer boundaries for certain model setups, 
for these cases we applied a larger physical grid size.

\subsection{Initial and boundary conditions}
\label{sec:icbc}
The initial conditions for our problem are as follows.
The jet is injected into an ambient medium of density $\rho_{\rm ism}$ and pressure $P_{\rm ism}$.
This gas is initially at rest and may carry a magnetic field $B_{\rm ism}$.
For all simulations, we assume a purely poloidal magnetic field in vertical direction, $B_{\rm ism} = B_y$.

Simulations are fundamentally governed by the boundary conditions.
In our setup, the jet is injected into the domain from a specialized jet nozzle of radius $\Delta x = 1$ with a 
velocity $v_{\rm jet}$.
For the magnetic field of the jet injection, we assume that this is equal to the magnetic field in the ambient 
medium, $B_{\rm jet} = B_{\rm ism}$.
We apply different jet nozzles providing jets that propagate substantially different flow channels.
The setup for these nozzles is discussed in detail in the next section.

The outer boundary conditions are {\em outflow} while along the $x$-axis we apply {\em reflective} 
boundary conditions outside the nozzle.

The injection properties together with the properties of the ambient medium define the leading
(magneto)hydrodynamic characteristics of the jet.
These are expressed in the Mach numbers of the flow.
That is the internal (sonic) Mach number, $M_{\rm s,int} = v_{\rm jet} / c_{\rm s, jet}$, 
with the sound speed in the jet material, $c_{\rm s, jet}$,
and the external Mach number $M_{\rm s, ext} = v_{\rm jet} / c_{\rm s, ism}$,
with the sound speed in the jet material, $c_{\rm s, ism}$.

In magnetized flows, the Alfv\'enic Mach numbers play the dominant role.
Similarly, we define internal Alfv\'en Mach number, $M_{\rm A,int} = v_{\rm jet} / v_{\rm A, jet}$, 
with the sound speed in the jet material, $v_{\rm A, jet} = B_{\rm p, jet} /  \sqrt{\rho_{\rm jet}}$.
The external Alfv\'en Mach number $M_{\rm A, ext} = v_{\rm jet} / v_{\rm A, ism}$ is defined via
the poloidal Alfv\'en speed in the ambient material, 
$v_{\rm A, ism} = B_{\rm p, ism} / \sqrt{\rho_{\rm ism}}$ \footnote{Note that the
factor $1\sqrt{4\pi}$ is absorbed in the definition for the magnetic field definition in PLUTO.}.

In order to investigate the curved jet structures, 
we mainly considered two general setups, that is the one of a precessing jet and the one of an orbiting jet
(see Table~\ref{tbl:box_sizes}). 
For the precessing jet nozzle we investigate five different initial deflection angles 
($\alpha_{\rm max} = 10, 30, 45, 60$ and $75\degr$).
For the orbiting jet nozzle we consider different separations and orbital periods, 
such as $(P_{\rm orb},a)$ is $(10,5)$, $(25,10)$ and $(25,5)$.

In order to investigate the effect of an interstellar magnetic, we re-run the hydrodynamic simulations 
{\em pj30deg} and {\em ojP25a10} applying a vertical magnetic field, initially
aligned with the jet, for field strengths considering a plasma beta $\beta=100,10$ and $1$.

We further detail the setup for the jet nozzles in the subsequent sections.

\begin{figure*}
\centering
\includegraphics[width=0.8\textwidth]{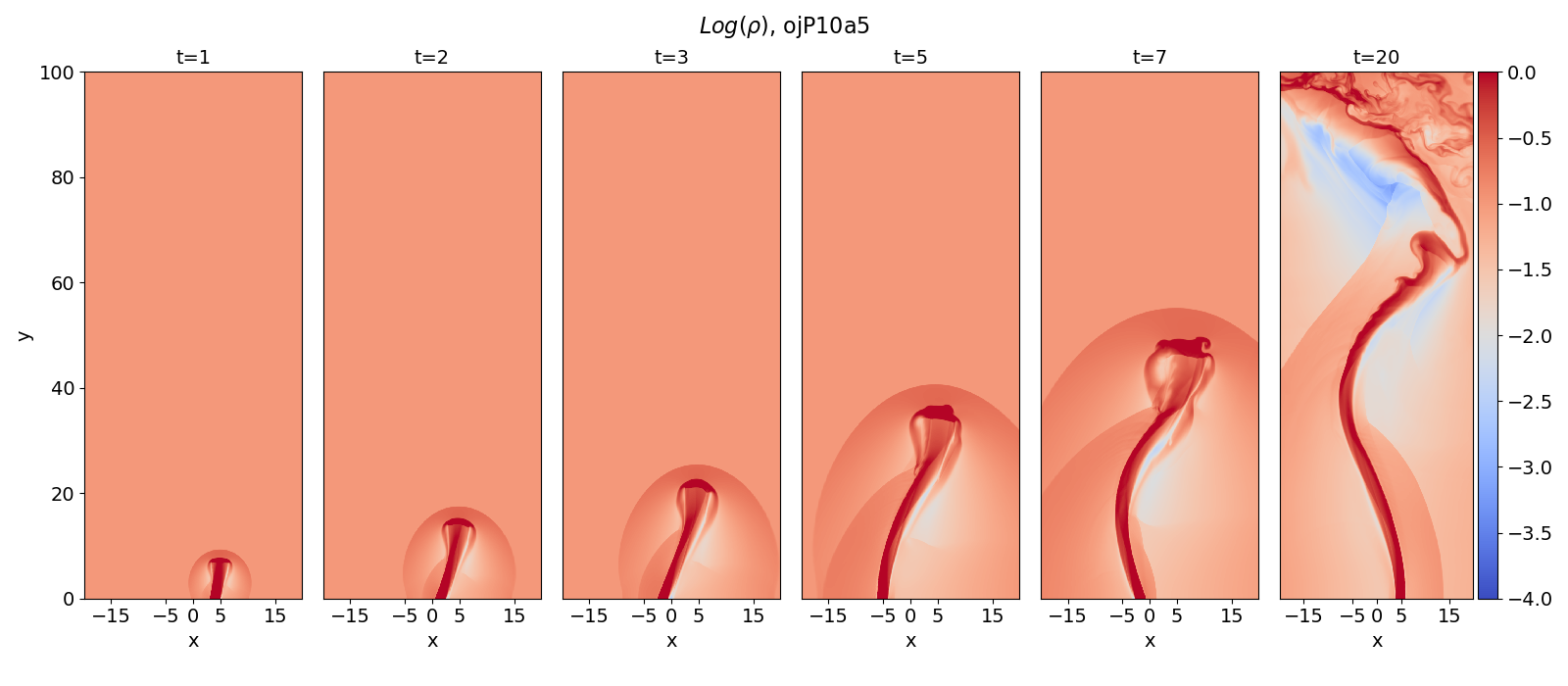}
\caption{Jet propagation from an orbiting jet source for run {\em ojP10a5}. 
Shown is the density evolution for times $t=1,2,3,5,7,20$.   }
\label{fig:sim1_rho-evol}
\end{figure*}

\begin{figure*}
\centering
\includegraphics[width=0.8\textwidth]{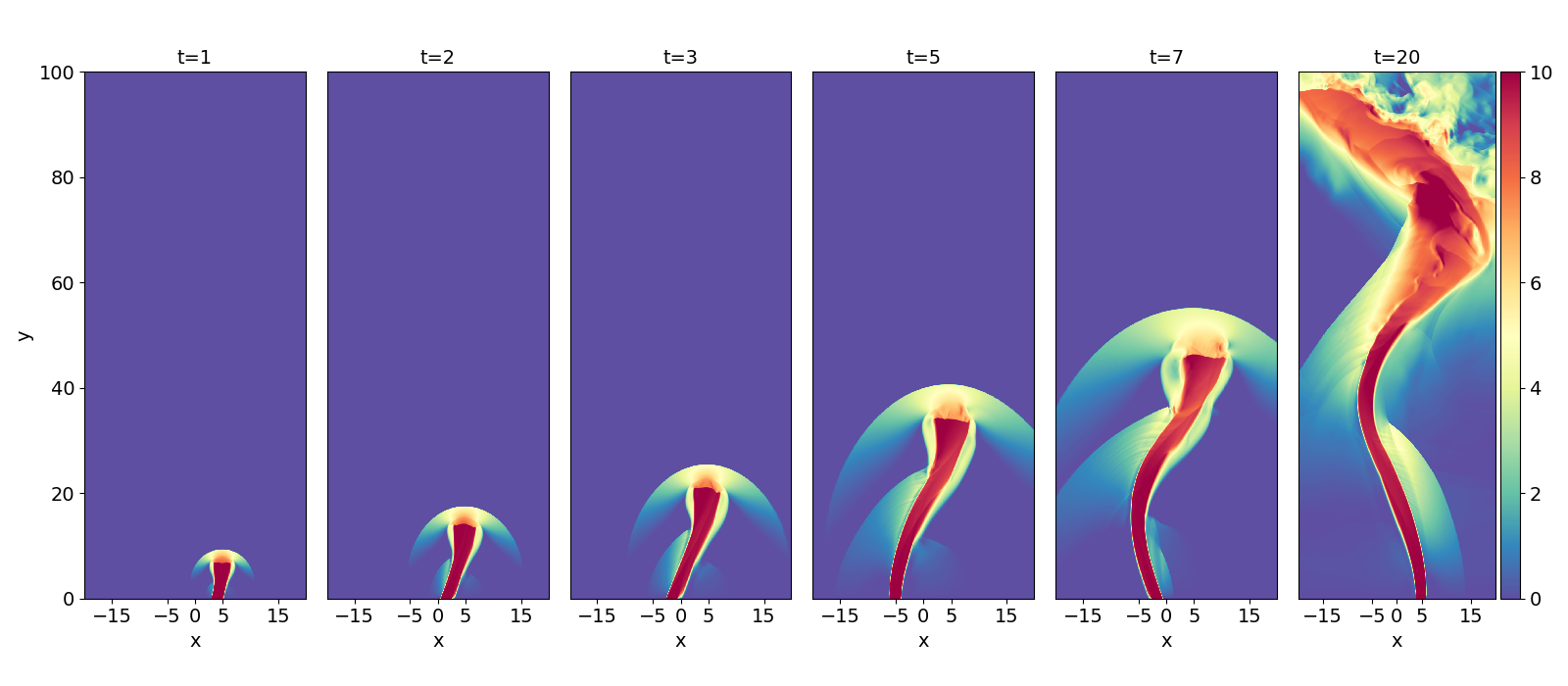}
\caption{Jet propagation from an orbiting jet source for run {\em ojP10a5}. 
Shown is the vertical speed for times $t=1,2,3,5,7,20$.   }
\label{fig:sim1_vx2-evol}
\end{figure*}

\begin{figure*}
\centering
\includegraphics[width=0.8\textwidth]{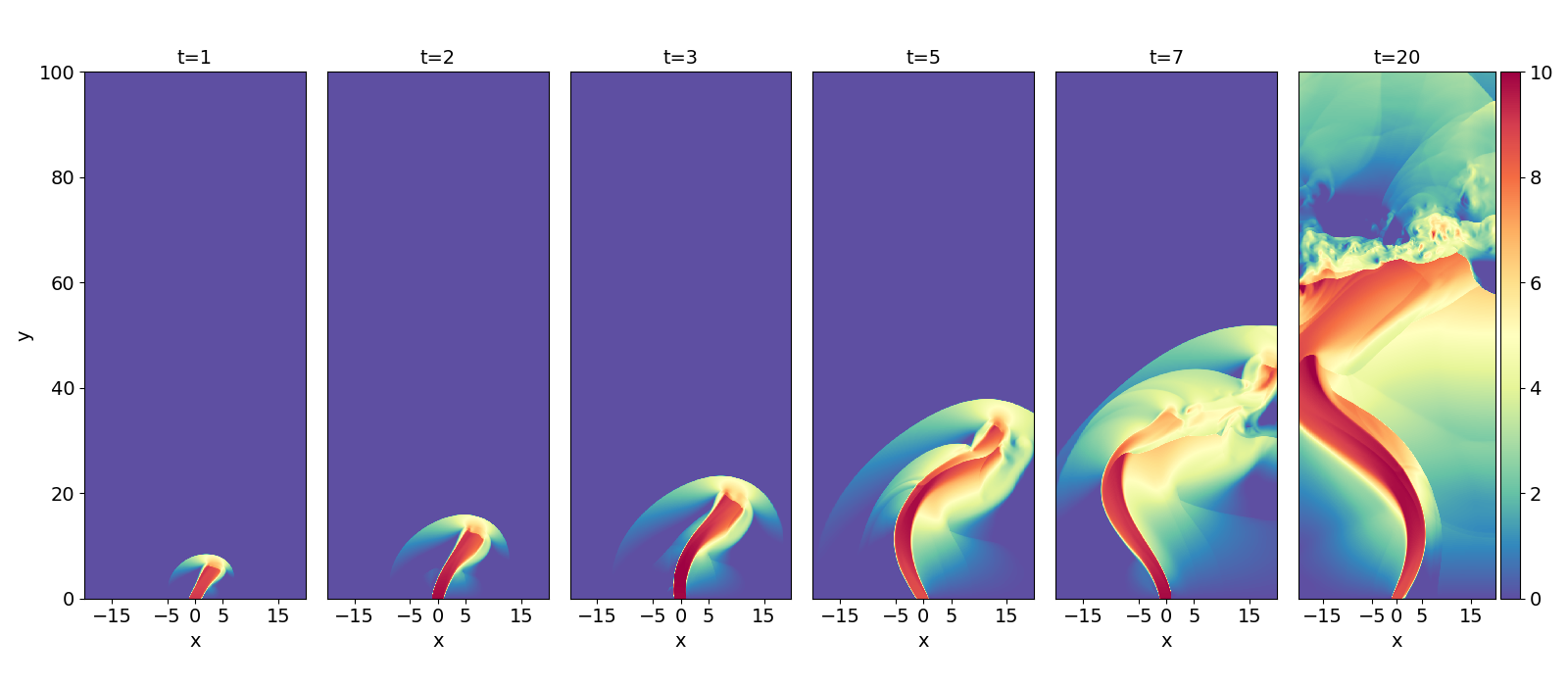}
\caption{Jet propagation from a precessing jet source for run {\em pj30deg}. 
Shown is the vertical speed for times $t=1,2,3,5,7,20$.  }
\label{fig:sim3_vx2-evol}
\end{figure*}

\section{Jet nozzles}
\label{sec:nozzles}
The definition of the jet nozzle is the fundamental boundary condition in our simulation setup.
It defines the injection properties of the jet and, thus, governs the path of the jet and 
also its interaction with the ambient gas, together with the properties of the ambient medium.

\subsection{Spatially fixed nozzles}
A spatially fixed jet nozzle is the standard setup that is used in almost all of the literature, starting with
seminal papers (see e.g. \citealt{1993ApJ...413..198S}) and is applied also for many recent simulations 
\citep{2005SSRv..121...21M, 2007ApJ...668.1028B, 2008A&A...486..663K, 2009ApJ...695.1376C, 2019ApJ...883..160S}. 
Spatially fixed nozzles that prescribe a time-variation of the jet injection (density, velocity) 
lead to pulsed jets \citep{1993ApJ...413..210S}.
This setup may provide periodic ejection of jet knots that are similar to observational features.
The setup, however, does not explain a mechanism that leads to a time-dependent ejection.
Recent approaches were applied to find periodic ejection of knots by a time-dependent disk dynamo mechanism 
that modifies the mass injection from the disk into the outflow by a variation of the disk magnetic flux 
\citep{Stepanovs2014}.

Applying multiple jet nozzles that are directed in different directions, colliding jets could be investigated 
\citep{2006ApJ...646.1059C}.
Also, the case of jets affected by side winds was investigated \citep{2004ApJ...616..988L}.

We will not further investigate a spatially fixed nozzle for injection but will apply jet nozzles that move in space.

\subsection{An orbiting nozzle}
\label{sec:nozzle_orb}
The model setup of an orbiting jet source is interesting for a number of astronomical sources.
Stars can be born as binaries and to investigate a binary setup for a protostellar yet is therefore interesting 
(see \citealt{Fendt1998} for an extensive discussion).
Jet launching conditions such as a strong magnetic field and an accretion disk can be present also in cataclysmic 
variables and high mass (compact star) binaries, and the question arises why not so many of these systems have jets.
There is also strong indication that some extragalactic jets may also be launched from supermassive black hole 
binary systems \citep{1980Natur.287..307B, 2013A&A...557A..85R, 2014MNRAS.437...32W, 
2017NatAs...1..727K, 2019ApJ...873...11J}.
If the jet is launched from one of the black hole system, its orbital motion may play a role.

With our model setup for an orbiting jet nozzle, we want to investigate how the motion of the orbiting jet source 
does affect the jet propagation on larger scales, and whether these effects can possibly be observed.
We also want to investigate whether there exist critical system parameters that prohibit large scale jet motion.

For our numerical setup in 2D Cartesian coordinates we approximate the orbital motion of the jet source as follows.
The jet nozzle with radius $\Delta x \equiv 1$ is located at $x_{\rm nzz}(t)$.
The motion of the jet nozzle is described by a sin function,
$x_{\rm nzz}(t) = a \cos(\Omega t)$, 
where $a$ is the binary separation, and $\Omega$ is defined by the orbital period 
$P_{\rm o} = 2\pi /\Omega_{\rm o}$.
Material of density $\rho_{\rm jet}$ is injected with speed $v_{\rm jet}$ from all ghost cells a position $x$
with $x_{\rm nzz}(t) - \Delta x < x < x_{\rm nzz}(t) + \Delta x$.

In code units $\Delta x = 1$.
Thus, the choice of the jet radius also determines the length scale of the system. 
The dynamical timescale is computed from $t_{\rm dyn} = 1 / v_{\rm jet}$.
This becomes essential when we re-scale from code units to astrophysical units.

Obviously, the strength of the effects that we want to compute depends directly
on the choice of the parameters $a$, $P_{\rm o}$, and $v_{\rm jet}$. 
In our simulations, we have applied a parameter choice that demonstrates these effects most clearly.
We may relate these simulation parameters to astrophysical units, even to different
astrophysical sources, however, it is clear that a perfect match which nature cannot be
reached.

In order to demonstrate the physical processes most clearly,
we choose $a = 5 \Delta x$.
As said, a suitable scaling of these code units depends on the astrophysics of the jet source.
For young stars, the following scaling seems plausible,
$\Delta x = 10$\,au, thus $a = 50$\,au for our simulation model.

The timescale in code units is thus $t_{\rm dyn} = 0.1$ for a jet speed of  $v_{\rm jet} = 10$.
For the orbiting jet sources, the orbital period chosen is $P = 10$ corresponding to 100 dynamical timescales.
The astrophysical timescale of jet propagation for this choice is 
$t_{\rm dyn} = 10\,\rm au / 50\,\rm km/s = 0.94 \,\rm yrs$ for a jet speed of $v_{\rm jet} = 50 \rm km/s$, or
\begin{equation}
 t_{\rm dyn}= 1.6 {\rm yrs} \left(\frac{R_{\rm jet}}{10\,\rm au}\right)  \left(\frac{v_{\rm jet}}{30\,{\rm km/s}}\right)^{-1},
\end{equation}
which could then be connected to the orbital period $P = 100 t_{\rm dyn}= 160$~yrs.
From this we calculate a total stellar mass of
\begin{equation}
 M_{\star}= 4.7 M_\odot \left(\frac{a}{50 \rm au}\right)^3  \left(\frac{P_o}{160\,\rm yr}\right)^{-2}.
\end{equation}
Note that $M_{\star} \propto R_{\rm jet} v_{\rm jet}^{2}$, considering the numerical scaling relations applied.

Considering symbiotic stars, there is common agreement that the compact star is a white dwarf 
below the Chandrasekhar limit. 
The stellar separations can fall within a wide range, from a few $\rm au$ to $\approx 10~\rm au$.
The geometrical size of these systems is definitely smaller than for a protostellar binary.
Here, for our astrophysical scaling considering the same set of normalized simulation parameters,
we find for $a = 5 \Delta x$ and $a \approx 4 \rm au$ a jet radius of $\Delta x = 0.8~\rm au$.

Thus, the dynamical timescale (again in code units $t_{\rm dyn} = 0.1$) would be
\begin{equation}
    t_{\rm dyn}= 14{\rm days} \left(\frac{R_{\rm jet}}{0.8\,\rm au}\right)  \left(\frac{v_{\rm jet}}{100\,{\rm km/s}}\right)^{-1}
\end{equation}
Assuming again an orbiting period of $P = 100$, corresponding to 3.8~yrs, the total stellar mass would be
\begin{equation}
    M_{\star}= 4.4\,M_\odot \left(\frac{a}{4\,\rm au}\right)^3  \left(\frac{P_o}{3.8\,\rm yrs}\right)^{-2},
\end{equation}
which is consistent with our assumption for symbiotic stars.

We emphasize that we do not intend to fit certain systems.
The previous estimates do only intend to show that our simulations results can be brought in rough agreement with
the different sources.
In particular, the jet speed applied is underestimated, which in turn emphasizes the effects of the lateral
motion we want to demonstrate.

We note that in our simulations, $v_{\rm jet} $ is typically quite larger than the prospective orbital speed of the jet source.
We may therefore neglect this orbital speed and the corresponding centrifugal forces, in agreement with our
setup of neglecting gravity for our simulations (see, however, \citealt{2002ApJ...568..733M} who performed 3D simulations).

The results for the initial jet evolution are shown in Fig.~\ref{fig:sim1_rho-evol} for the density
and in Fig.~\ref{fig:sim1_vx2-evol} for the vertical velocity.
We clearly see a periodic jet bending.

The curvature and wavelength of the jet bending depend on the orbital period, the jet speed and the the stellar separation.
For the parameters applied, we find the amplitude for the curved jet motion being initially similar to the
orbital radius.
This may be somewhat expected.
For the second jet {"}wave{"} the amplitude increases to about two orbital radii.
The wavelength of the curved jet propagation is about ten orbital radii.

For a {\em bipolar} jet source, the evolving flow structure is mirror symmetric with respect to the mid-plane,
respectively the orbital plane.
Thus, a jet emitted from an orbiting jet nozzle shows a {\em C-shaped structure} on larger scales.
However, with time also the conditions with in the jet channel and the environment change, and the
question arises how long such a periodic structure will survive on the long term.
We will investigate this question in Sect.~\ref{sec:channel}.

\subsection{A {"}precessing{"} nozzle}
\label{sec:nozzle_prc}
Indication for jet precession has been observed in a number of sources of different classes.
Precession is expected for jet sources in binary systems in cases that the rotational and orbital axes are misaligned.

Since precession is most probably a result of the binarity of the jet source, 
we would expect effects due to the orbital motion as well for the precessing nozzles.
However, for the sake of simplicity we do not investigate the combined action of binary effects 
here and prefer to disentangle the effects caused by {\em either} orbital motion {\em or} jet precession.
We also note that the precession timescale is usually many orbital timescales.
We thus expect the two effects to work on different time scales and can treat them separately.
In principle, our setup would allow us to put a precessing nozzle on an orbit and measure the combined effect – prescribing a 
similar time scale, $P_{\rm orb} \simeq P_{\rm prc}$. 
However, we expect that from the results of simulating the combined effects, it would be almost impossible to disentangle 
the origin of certain features.

The precession period ($P_{\rm prec}$) can be related to the orbital parameters as
\begin{equation}
 P_{\rm prec} = \frac{64}{15}\pi \frac{M_{\rm p}}{M_{\rm s}}\left( \frac{a}{R} \right)^3 \sqrt{\frac{R^3}{G M_{\rm p}} } \frac{(1-e^2)^{3/2}}{\cos i}  
\end{equation}
(see e.g. \citealt{Beltran2016}), that can also be applied to wide binaries \citep{1995MNRAS.274..987P}. 
Here, $M_{\rm p}$ and $M_{\rm s}$ are the mass of primary and secondary stars respectively,  
$R$ is the radius of the disk, $a$ is a separation between binary stars, $e$ is the eccentricity of the orbit, 
and $i$ is the inclination of the binary orbit with respect to the plane of the disk.

For close binaries, the spin axes and the orbital plane of the sources are expected to be aligned via the magnetic forces.
However, (super)outbursts may cause a certain degree of misalignment. 
Superhumps, defined by their light curve and are caused by super-outbursts, arise rapidly but fade on time scales longer than the orbital period. 
Therefore, the relation between the superhump period $P_{\rm sh}$ and the respective orbital period $P_{\rm orb}$ is 
$P_{\rm prec} = P_{\rm orb} P_{\rm sh}/( P_{\rm sh} - P_{\rm orb})$ \citep{Warner95}.

We approximate the precession motion of the jet source in our 2D Cartesian setup as follows.
The jet nozzle with radius $\Delta x \equiv 1$ is now located at the radius $x_{\rm jet}(t) = 0$.
The jet is injected in a prescribed direction $\alpha_{\rm jet}$ (measured from the rotational axis) 
that varies in time, $\alpha_{\rm jet}(t) = \alpha_{\rm jet, max} \cos(\Omega_{\rm p} t)$, between the angles 
of maximum deflection, $\alpha_{\rm jet, min}$ and $\alpha_{\rm jet, max}$.
$P_{\rm prec} = 2 \pi /\Omega_{\rm p}$ and we consider $P_{\rm prec} = 10$ in code units for all precessing nozzle simulations.

The jet speed $v_{\rm jet}$ remains constant in time, however the velocity components of the injected jet
vary over time as the direction of the nozzle changes,
$v_{x,\rm jet} = v_{\rm jet} \sin(\alpha_{\rm jet})$, and $v_{y,\rm jet} = v_{\rm jet} \cos(\alpha_{\rm jet})$.

In Figure~\ref{fig:sim3_vx2-evol} we show the initial time evolution of the system, applying 
$P_{\rm p} = 2\pi / \Omega_{\rm p} = 10$, 
$v_{\rm jet} = 10$, and $\alpha_{\rm jet, max} = 30\degr$.
The same time scaling  as it is described for the orbiting jet source applies, thus $t_{\rm dyn} = 1 / v_{\rm jet}$.

We immediately see the {\em essential difference} to the model of an orbiting jet source, such that for a 
{\em bipolar} jet,
the evolving flow structure is {\em mirror symmetric} with respect to the origin.
Thus, a jet emitted from a precessing nozzle shows a {\em S-shaped structure} on larger scales.

\begin{figure}
\centering
\includegraphics[width=\linewidth]{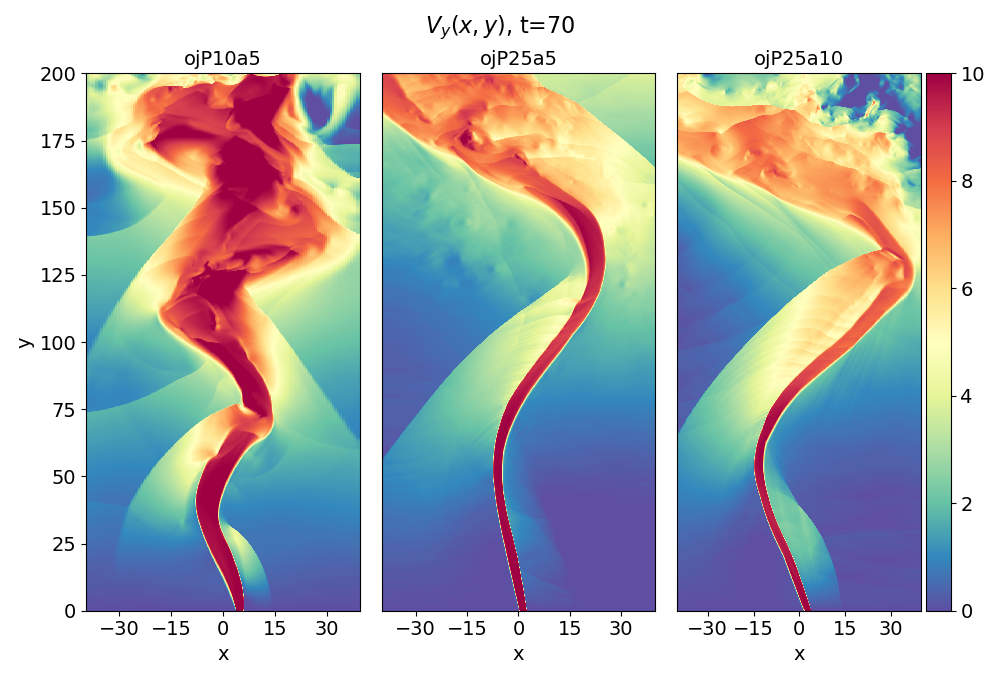}
\caption{Hydrodynamic jet evolution for different orbital parameters, shown at time $t=70$,
for combinations (from left to right) of periods and separations $(P=10, a=5)$, $(P=25, a=5)$ and $(P=25, a=10)$,
corresponding to simulation runs {\em ojP10a5}, {\em ojP25a5} and {\em ojP25a10}, respectively.     }
\label{fig:t70_dif_bina_param}
\end{figure}

\begin{figure}
\centering
\includegraphics[width=\linewidth]{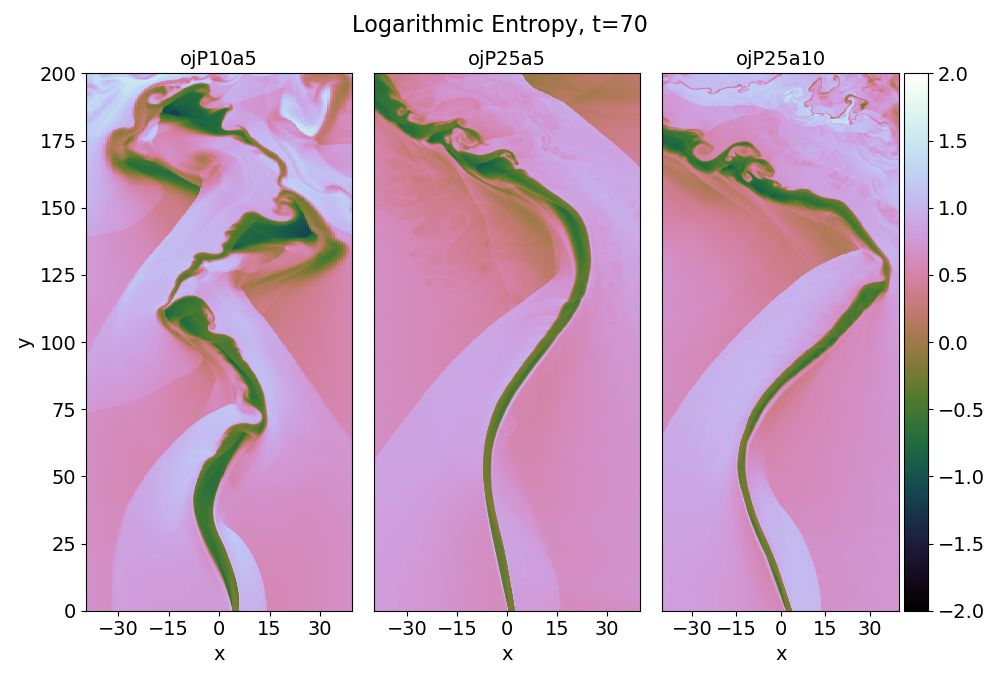}
\caption{Same as Fig. \ref{fig:t70_dif_bina_param}, 
but for the measure of entropy $K \equiv P / \rho^{\gamma}$ (in log scale).    }
\label{fig:t70_dif_paramentropy}
\end{figure}

\begin{figure*}
\centering
\includegraphics[width=0.8\textwidth]{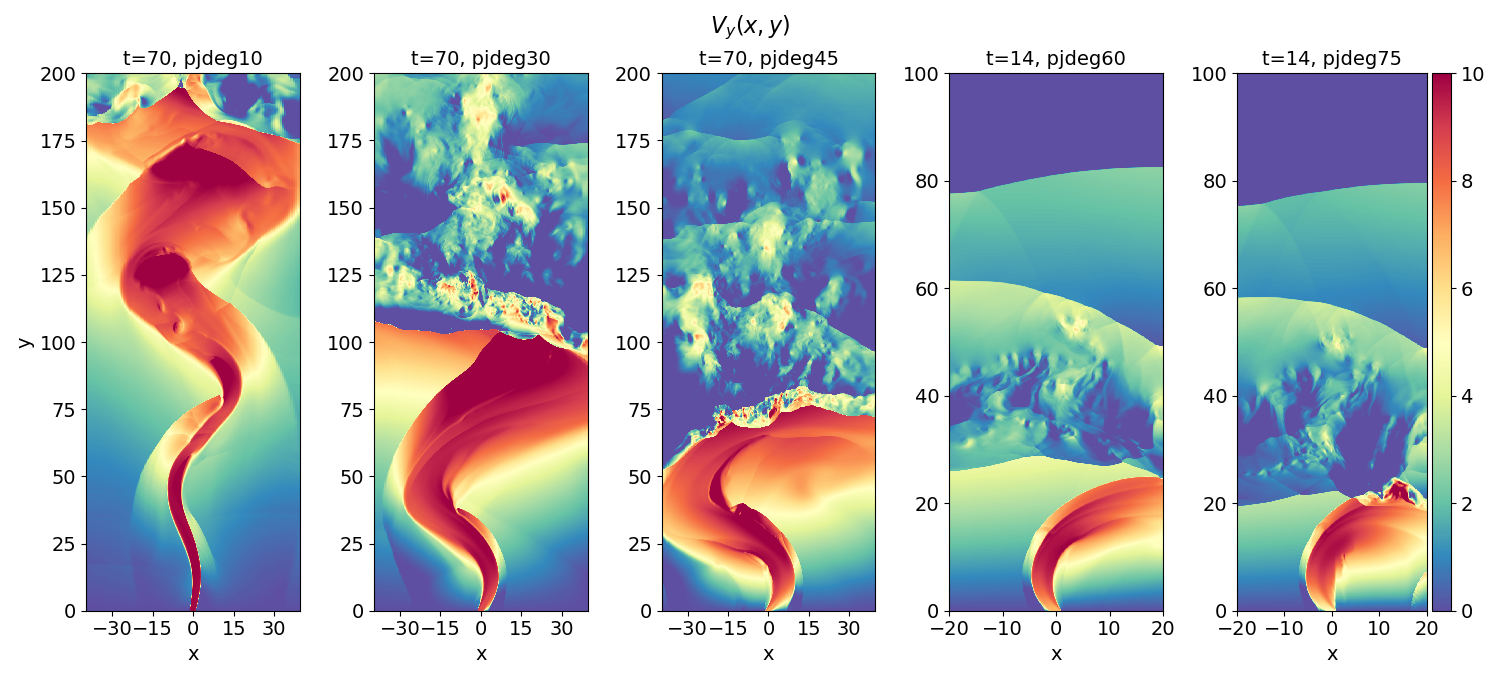}
\caption{Hydrodynamic evolution for jets of different precession angles, $\alpha = 10, 30, 45, 60, 75^{\circ}$
(from left to right), corresponding to simulation runs 
{\em pjdeg10}, {\em pjdeg30}, {\em pjdeg45}, {\em pjdeg60} and {\em pjdeg75},
at time steps $t= 70, 70, 70, 14, 14$, respectively.
Note that a large injection angle $\alpha > 45^{\circ}$ for the nozzle results in a jet propagation that 
fails to reach large distance.    }
\label{fig:t70_dif_prec_param}
\end{figure*}

\section{C-shaped jets from orbiting jet sources}
\label{sec:c-shape}
We now discuss our simulation results on intermediate time scales for the setup of an orbiting jet nozzle 
(see Sect.~\ref{sec:nozzle_orb}).
Figure~\ref{fig:t70_dif_bina_param} shows the evolution of the jet beam at the same time for different
orbital parameters, but at the same evolutionary time step $t=70$.
All simulations were initiated with the same initial conditions for the ambient medium, $\rho_{\rm ism}$, $P_{\rm ism}$, 
and run with the same injection parameters for the jet $\rho_{J}$, $P_{J}$, $v_{J}$.
However, different orbital periods and orbital separations were applied.
Naturally, we would expect a correspondence between the jet stability on larger scales with these
parameters that determine the {"}disturbance{"} of a straight jet motion.

The disturbed jet propagation is characterized by its {\em curvature}, that is, a combination of the
wavelength and the amplitude of the perturbed jet geometry.
Obviously, the jet curvature is determined by the orbital separation
(defining the amplitude), and the time scales for the orbital motion and the jet speed.

For the examples shown, we find that the wavelength $\lambda$ of the jet propagation seems to be determined by the 
orbital period rather than by the orbital separation
(about $\lambda \simeq 40$ for $P=10$, and $\lambda \simeq =60$ for $P=25$).
On the other hand, the amplitude $A$ of the jet bend seems determined by the orbital separation 
(about $A \simeq 10$ for $a=5$, and $A \simeq 15$ for $a=10$).

Another signature that can be easily derived from the simulation result (and can be compared to
observed sources) is the length of the jet.
We clearly see that the simulations considering a wide orbital radius $a = 10$, become unstable the earliest.
In particular, these jets injected do not survive long and do hardly reach the upper boundary.
Typically, the jet flow is smooth and well defined for at least one bend, after which some of the flows break up, 
depending on the orbital parameters.

From the astrophysical point of view, these jets are shorter and do not penetrate the ambient medium for
long.
The jet energy is converted into thermal energy more efficiently due to the large effective surface 
of the jet.
Note that the kinetic energy flux of all these jets is the same.

We also find a clear relation between the wavelength and the amplitude of the jet bending and the orbital 
parameters.
The shorter the orbital period (for the same stellar separation) the shorter the wavelength.
Altogether, these seem to be too strongly disturbed and do not survive long when propagating.
Astrophysically, such systems correspond to binaries containing a relatively massive primary (the jet being 
ejected from the secondary), for example a massive compact primary with a low mass stellar companion.
Here, we do not consider here gravitational torques (see for example \citealt{2015ApJ...814..113S,2022ApJ...925..161S}), 
but alone from the orbital motion of the jet source, we may conclude that such binaries may not eject stable 
narrow jet streams propagating much longer than the respective orbital separation.
This is in nice agreement with long-term 3D simulations of MHD jet launching in binary systems
which find a critical inclination between accretion disk and orbital plane of about $10-20\degr$, 
resulting in a precession cone of the jet of $8\degr$ \citep{2018ApJ...861...11S}.

Jets originating from binaries with small separation and short orbital period, however, are able to travel as
turbulent beams to relatively long distances.
We understand that this becomes possible, as due to the small orbital separation the jet injections over time
widen the funnel that was already drilled by the previous injections, resulting in a clean and broad jet funnel.
In contrast, for jets originating from binaries with wider separation, each jet injection (over time) practically 
bores its own funnel into the ambient medium.
Thus, instead of a single wide funnel the jet injection produces its own small funnel.
These small funnels are, however, terminated when the jet injection continues on its orbit, and ejects in
new directions.
 
In Fig.~\ref{fig:t70_dif_paramentropy} we show the entropy maps for simulation runs {\em ojP10a5}, {\em ojP25a5}, 
and {\em ojP25a10}.
Entropy can show the origin of the gas as it labels the gas temperature, thus showing whether we see cool (dense)
gas that was injected by the jet nozzle, or hot gas (of the originally low density gas of the ambient medium).

We see that at the turning points of the jet motion the matter is pushed further, and a shock front is formed. 
Here, the kinetic energy is transferred to the shocked gas, heating it (to larger entropy). 
The backside of the shock stays cool. 
Across the shock front, the entropy increases, especially in the turning points of the jet mentioned above, while
the inner jet stream has low entropy. 
Whenever a shock is produced, it disrupts the continuity of the flow, causing a turbulent flow pattern. 
However, at these turning points, the flow accumulates into filament-like structures.
These structures depend somewhat on the shock properties.
The first panel of Fig.~\ref{fig:t70_dif_paramentropy} clearly demonstrates the generation of turbulence across 
the turning points of the jet.

We finally note that these are 2D simulations and that the bent or disturbed jet structure will certainly give rise
to further jet instabilities, such as kink modes or shear instabilities.

The amplitude of the jet bending seems simply related to the orbital separation.
The larger the orbital separation the larger the amplitude.
However, for the same amplitude, the wavelength becomes larger for larger orbital periods.

A well-known example for a C-shaped jet structure is HH~211, for which binary components of different mass
are discussed that may cause the jet bending \citep{2009ApJ...699.1584L}.
Another C-shaped jet source is HH~212 (see e.g. \citealt{2015ApJ...805..186L,2020RMxAA..56...29N}), again the 
deviation from the linear motion is discussed in terms of a binary motion of the jet-ejection star.

\begin{figure*}
\centering
\includegraphics[width=0.8\textwidth]{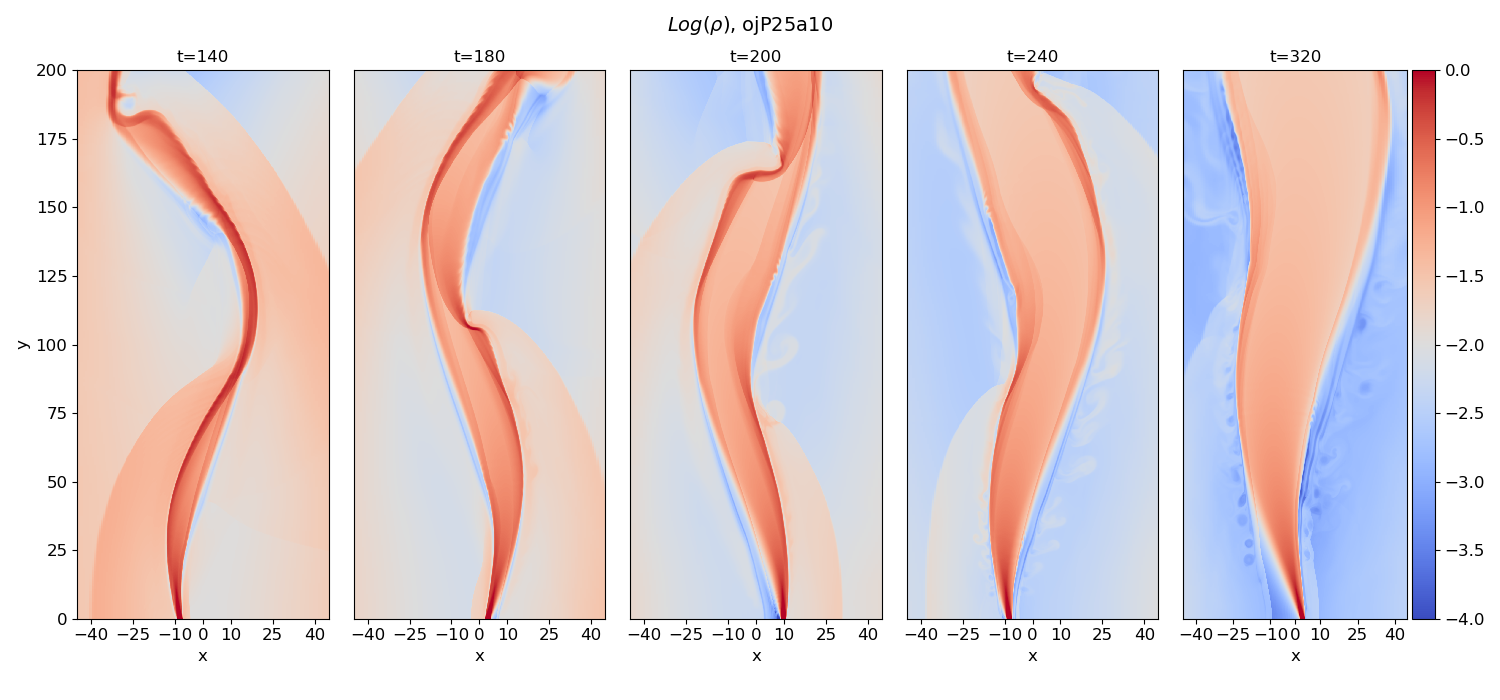}
\caption{Long term evolution for run {\em ojP25a10}. 
Shown is the density in log scale for time steps $t = 140$,  $t = 180$, $t = 200$, $t = 240$, and $t = 320$.}
\label{fig:widejet_rho1}
\end{figure*}

\begin{figure*}
\centering
\includegraphics[width=0.9\textwidth]{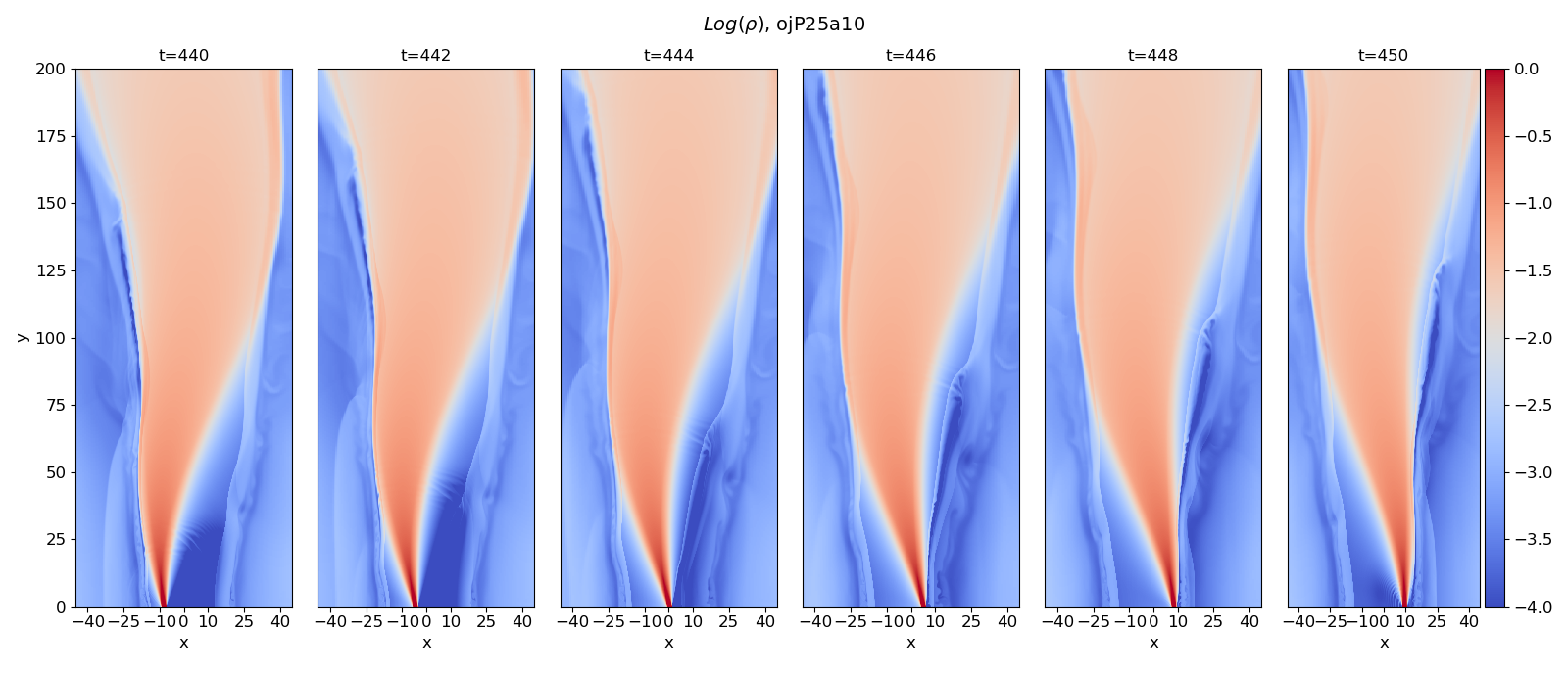}
\caption{Late-time evolution for simulation {\em ojP25a10} from $t=440$ to $t=450$.
Shown is the density in log scale for a sequence with high time resolution, $\Delta t = 2$.}
\label{fig:widejet_rho2}
\end{figure*}

\section{S-shaped jets from precessing jet sources}
\label{sec:s-shape}
We now discuss our simulation results on intermediate time scales for the setup of a precessing jet nozzle 
(see Sect.~\ref{sec:nozzle_prc}).
Figure~\ref{fig:t70_dif_prec_param} show the evolution of the jet beam at the same time for different
precession angles of the jet nozzle at time $t=70$ for the longitudinal velocity distribution.
 
We clearly see how the opening angle of the precession cone affects the jet propagation on longer scales.
Jet beams that are ejected along a widely opened cone, do not propagate very far.
The critical angle is between $10\degr$ and $20\degr$, but of course, also depends on the further jet
kinematics.
We explain this as follows.
For large precession angles the momentum that is injected into the jet in the direction of the axis of
the precession cone, is not sufficient for penetrating the ambient medium.
The latter is stirred up, as visible in the turbulent-looking structure and the shock(lets) that are moving
away from the jet-ISM interface.

These outflows may for example trigger a local collapse in the ISM.
However, their penetration depth is some 50-100 initial jet radii only.
The outflow moves along their outflow cone with almost their injection speed, but are quickly halted at the 
shock that develops at the very front of the outflow.  

For precession angles $< 45\degr$, at early times, the jet flow shows a {\em C-shape} geometry. 
As time progresses, the jet stream begins to turn into an {\em S-shaped} form.
For larger precession angles an {\em S-shaped} jet could not be produced.
As it does not propagate sufficiently long distance due only a short-reaching {\em C-shaped} jet appears, 
see Fig.~\ref{fig:t70_dif_prec_param}. 

Another interesting feature we note is the jet curvature increases with distance from the injection.
For example, simulation {\em pjdeg10} that applies a nozzle opening cone of only $2\times 10\degr$ evolves into
much more curved structure before the jet terminates. 
In fact, with each wiggle (there are three of them), the curvature radius decreases, from say 30 to 15 to 10.
In fact, the reason for the jet termination is exactly this decrease of the corkscrew pitch angle, that leads
to a constant momentum loss along the original jet direction.
In the end, the jet has lost too much of its momentum and cannot easily penetrate further.

When the jet curvature increases, material accumulates at the turning points of the jet. 

Along the jet stream, the jet opening cone expands on large distances.
We see an indication that jet knots are produced (in particular for simulation {\em pjdeg10}). 

On the other hand, the wiggling of the highly inclined jet {'}tail{'} introduces shock waves into the ambient
gas, best seen for {\em pjdeg60} and {\em pjdeg75} simulation runs seen in the velocity slices in 
Figure~\ref{fig:t70_dif_prec_param} (right).
The shock separation and shock speed are determined mostly by the precession time scale.
The shocked material is moving with a speed $v_{\rm s}\simeq 2$, defining a timescale between the shocks
of $\Delta t_{\rm s} = \Delta y /  v_{\rm s} \simeq 30/2 = 15$.
This timescale is in good agreement with the precession timescale $P_{\rm prec} = 10$.

Even when the jet propagation is limited in distance, more and more material is injected into
the jet channel. 
As it cannot propagate further, this material is accumulated and the jet head broadens.

Many protostellar jets are observed only one-sided, but there are prominent examples that
show a S-shaped propagation patter, such as L1157 \citep{2016A&A...593L...4P}.
Another example is MHO 2147 \citep{2022A&A...657A.110F} that also seems to follow the analytical
trajectories for precessing jets proposed by \citet{2002ApJ...568..733M}.
A further protostellar jet indicating on precession is DO~Tau \citep{2021A&A...650A..46E}.

For the symbiotic system R~Aqr jet components can be traced to up to $\sim 900$~au, which is quite long 
compared to the binary separation of $\sim15$ AU. 
The jet is surrounded by an outflow of a wide opening angle filled with low excitation gas. 
\citet{Melni2018} mention the origin of the gas emission can be cold stellar wind, especially from mass 
losing Mira variable and partly collimated jet. 
This observational findings could be nicely matched by our simulations such that this cocoon is filled by
shock gas at the locations when the wave of the jet stream has its turning points. 
This wide opening cone along the jet we see in particular in our long time step simulations. 
\citet{Melni2018} in particular mention the S-shaped jet in the inner part of the wide-outflow 
close to the stellar system which is exactly what we find in our long-term simulations (see below). 

Another example of a curved jet is CH~Cyg \citep{Karovska10}, another symbiotic star. 
Interestingly, this jet is changing its morphology over time between S-shaped and C-shaped.
Its shock front is propagating with <100km/s and is slowing down since 2001. 
The authors suggest that this jet can be produced by an episodically powered precessing jet or continuous 
precessing jet with occasional mass ejections or pulses due to the structure of the jet. 
CH~Cyg may thus be a good example of a jet where both the nozzle effects discussed above can play a 
similar role, and the jet structure that is visible is originally triggered by a time-variable feeding 
to the jet.
It would be thus interesting for a future simulation to simulate a precessing {\em and} orbiting nozzle
in order to find transitions in morphology.

We note that at this stage we cannot intend to fit certain courses as many important dynamical
parameter are observationally not yet determined. 
We are left with finding similarities in the jet propagation that may hint on the nature of 
the jet source, such as precession vs. orbital effects. 

\section{Long-term evolution: transition to a wide jet channel}
\label{sec:channel}
We now investigate how the jet propagation evolves on much longer time scales.
We have seen above that for intermediate time scales (and for particular system parameters) the jet evolves as a bent, 
wavy structure, similar to a rope that is perturbed at one end.
The wave pattern propagates until it terminates at the bow shock.

This picture changes when we consider much longer time scales.
In Figure~\ref{fig:widejet_rho1} we show the evolution of the same injection model that we considered above (see 
Fig.~\ref{fig:sim1_rho-evol}), but for times $t=140$ to $t=320$.
We find that the wave pattern broadens such that the wavy channel of the jet flow becomes wider and wider.
However, the jet flow is still oscillating.

At some point in time, the lateral motion of the oscillating jet has produced a conical-shaped outflow channel 
with a geometry that has become rather steady in time.
This happens roughly for times $t > 400$ (see Fig.~\ref{fig:widejet_rho2}).
The half opening angle of the cone is about 15\degr.

The width of the cone at the injection point is about the orbital diameter.
The jet nozzle orbits within this cone, while the injected jet material is distributed over
the whole width of the conical outflow.
This can be seen in Fig.~\ref{fig:widejet_rho2} where we show the evolution in high time resolution.
The wave pattern of the jet motion is gone.
However, the boundaries between the wide jet flow and the ambient gas are not straight, but still show a kind of wavy 
structure, but this seems to be decoupled from the forcing perturbing the jet injection.
The orbital jet nozzle still governs the initial jet flow and imprints a periodic wiggling on it.
However, on more distant scales the lateral outflow seems decoupled from the motion of the nozzle.

\begin{figure*}
\centering
\includegraphics[width=0.8\textwidth]{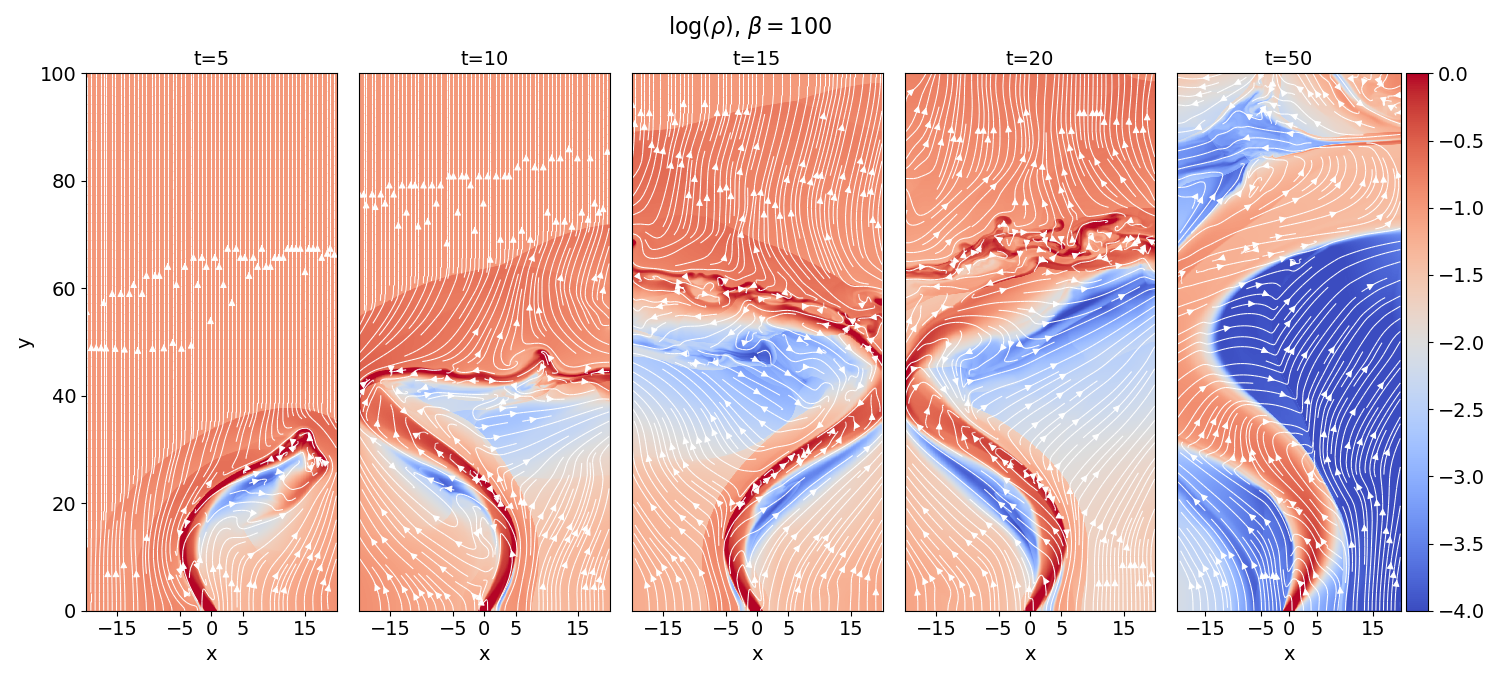}
\caption{Time evolution of the magnetized simulation pj30degB100 with a maximum precession angle of $\alpha_{\rm max}=30\degr$
and for $\beta=100$.
Shown is the density distribution (colors) and the magnetic field lines (white) for the time steps $ t = 5,10,15,20,50 $ (left to right). }
\label{fig:beta100_flines_rho}
\end{figure*}

\begin{figure*}
\centering
\includegraphics[width=0.8\textwidth]{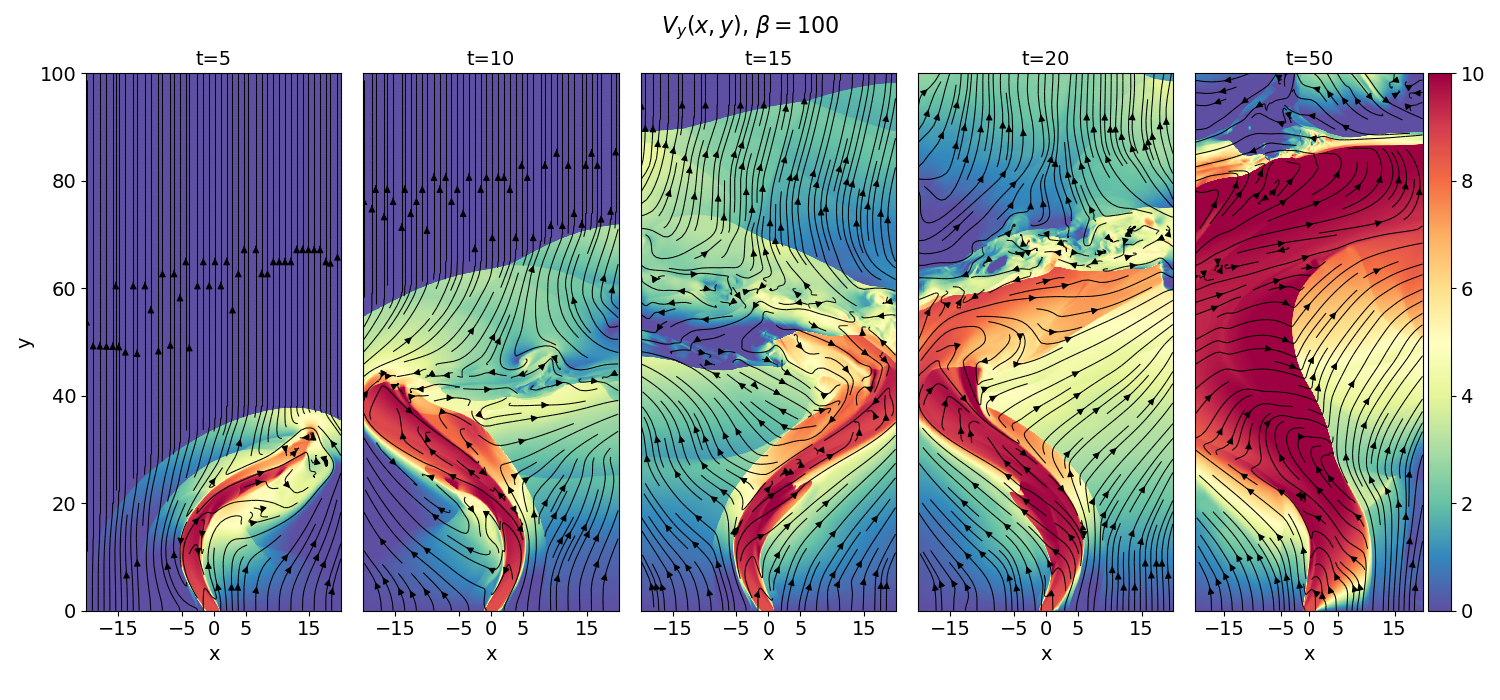}   
\includegraphics[width=0.8\textwidth]{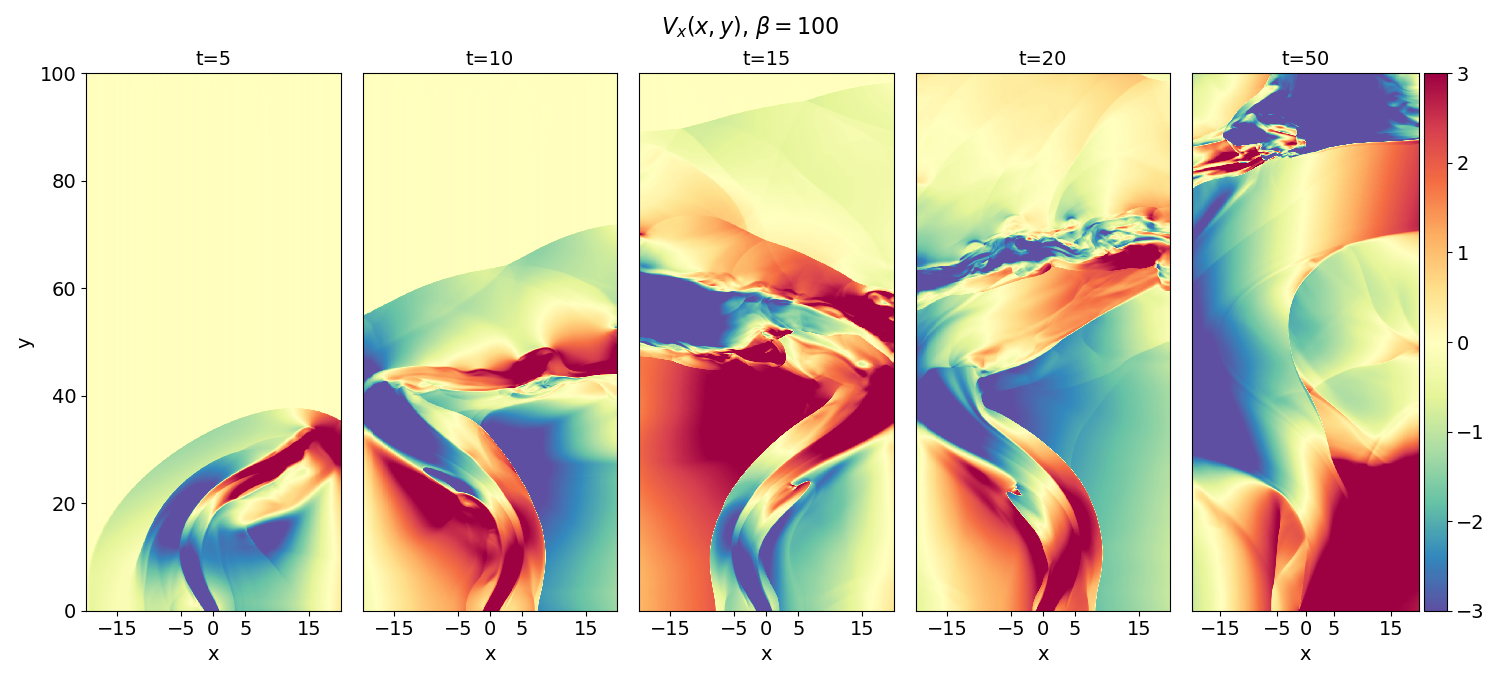}
\caption{Time evolution of the magnetized simulation {\em pj30degB100} with a maximum precession angle of $\alpha_{\rm max} = 30\degr$
and for $\beta = 100$.
Shown is the vertical (top) and horizontal (bottom) velocity distribution (colors) and the magnetic field lines (black) 
for the time steps $t=5,10,15,20,50$ (left to right). } 
\label{fig:beta100_flines_v}
\end{figure*}

\begin{figure}
\centering
\includegraphics[width=\linewidth]{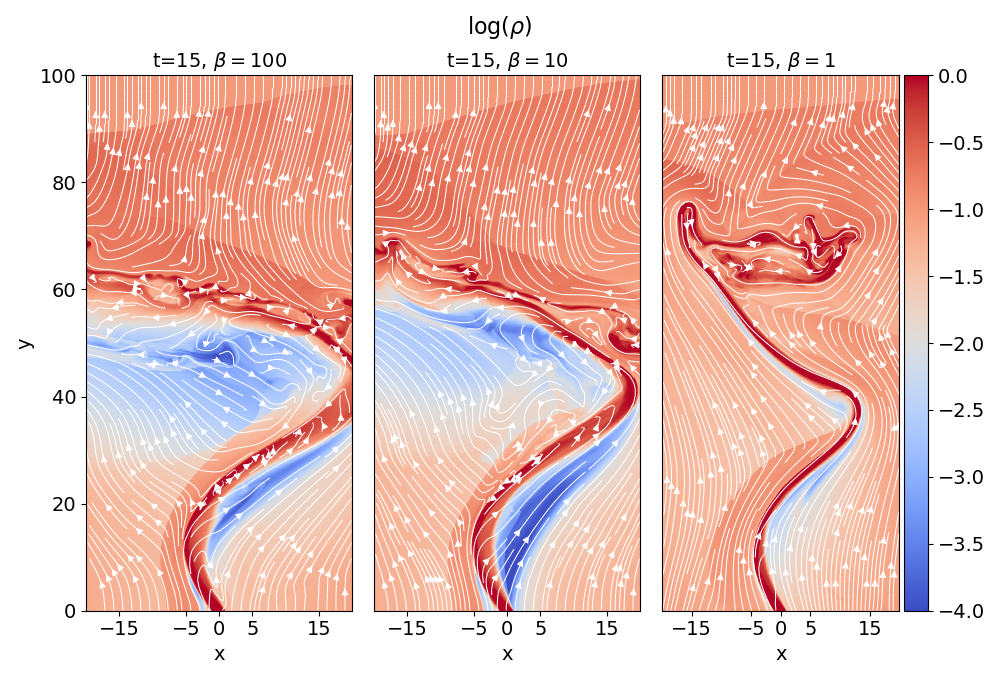}
\caption{Precessing jets injected into an ambient gas of different plasma beta.
Shown is the density and magnetic field lines at time $t=15$ for $\beta = 100, 10, 1$ from left to right, respectively.
The precessing angle of the jet nozzle is $\alpha_{\rm max}=30\degr$.}
\label{fig:betas_flines_rho}
\end{figure}

\section{Impact of the magnetic field}
\label{sec:magnfield}
Astrophysical jets are magnetized.
We know that jet launching is in general a magnetic phenomenon requiring a large-scale magnetic flux in the jet driving 
source
\citep{1977MNRAS.179..433B,1982MNRAS.199..883B, 2002ApJ...581..988C, 2015SSRv..191..441H}.
Also, the medium into which the jet is injected is magnetized.

An interesting question is the alignment of protostellar jets with the magnetic field in the ISM.
This still seems to be an open issue with various papers providing different answers
\citep{2011ApJ...743...54T, 2013ApJ...768..159H,2019FrASS...6...54P,2019MNRAS.485.4667H}.
In our treatment, for simplicity, we adopt a magnetic field geometry that is aligned with the average jet injection axis.
This allows disentangling the magnetic effects most clearly.

\subsection{The question of a toroidal field}
It is clear that magnetized jet typically carries also a toroidal magnetic field. 
This comes as a natural consequence of MHD models, as the inertia of the jet material induces a toroidal field 
from the poloidal component \citep{1982MNRAS.199..883B, 1985PASJ...37..515U, 2016ApJ...825...14S}.
The toroidal field plays an important role for jet collimation and stability, however, it is likely impossible to 
implement it in a 2D Cartesian setup in a consistent way.

A possible approach may be to inject a $B_z$ component that is anti-aligned left and right from the jet axis.
However, this approach is yet untested in the literature and its benefit is maybe little.
Jet collimation and stability are not of interest to the present study.
What we neglect here, is thus mainly the magnetic pressure by the 3rd field component.
Tension forces by the 3rd field component cannot be derived are as its curvature cannot be considered 
in a 2D slab approach.
Overall, the main effect of the 3rd field component would be an enhanced overall jet pressure, which could,
however,  mimicked by applying an increased gas pressure.

We note that without considering a toroidal magnetic field (thus neglecting poloidal electric currents), 
we cannot investigate possible jet bending by Lorentz forces, $\vec{F}_{\rm jet} \propto \vec{j}_{\rm jet} \times \vec{B}_{\rm ism}$,
as previously suggested by \citet{1990A&A...232...37M} or \citet{Fendt1998}.

\subsection{A magnetized precessing nozzle}
We have re-run the hydrodynamic simulations with a different ambient poloidal field strength, defined by the
plasma beta $\beta \equiv P / B^2$, where $P$ is the initial gas pressure (of jet and ambient gas)
and $B$ is the strength of a constant magnetic field in vertical direction (aligned with the mean direction 
of injection).

For this section, we have concentrated our discussion on the model setup of a {\em precessing} jet nozzle.
We have found that the main effects of the magnetic field on the curved jet motion in the simulations
for an orbiting nozzle are basically the same (see below and the Appendix). 

Assuming a constant magnetic field implies that the initial field is in fact force-free with no
magnetic pressure or magnetic tension forces imposed.
Any material motion perpendicular to the field direction will induce perpendicular field
components which then lead to Lorentz forces providing feedback to the original jet propagation.

Figure~\ref{fig:beta100_flines_rho} shows the time evolution of density and magnetic field 
for simulation pj30degB100.
Here, a maximum precession angle of $\alpha_{\rm max}=30\degr$ and a plasma-beta of  $\beta=100$ is applied.
We clearly see the magnetic field is perturbed by the jet motion. 
The dynamics of the wiggling jet is dominating the field distribution.
The field lines tend to follow the jet motion with exposing a similar curvature.

Similar behavior can be seen in the velocity distribution in Figure~\ref{fig:beta100_flines_v}.
Notably, the velocity distribution is substantially broader than the density distribution.
The reason is the material that has been entrained along the high-speed jet.
Obviously, this holds for both the vertical and horizontal velocity components.

In Figure~\ref{fig:betas_flines_rho} we compare examples of our simulations.
Here, a vertical field of different strengths was supplied to the pure hydro setup of simulation
\textit{pj30deg}.
The injection nozzle was assumed to precess with an opening cone of $30\degr$.
The plasma beta imposed initially is $\beta = 100, 10, 1$.

The simulations clearly demonstrate  the stabilizing effect of the magnetic field.
While for high plasma beta the wiggling jet dominates the magnetic field and thus wiggles the magnetic field
along with the time-dependent jet injection, this wiggling of the jet (also compare to the hydro case) and
the magnetic field is weaker for a low plasma beta.
As a consequence, the jet motion remains relatively undisturbed, and the stream can reach further distances 
from its injection site.

\begin{figure*}
\centering
\includegraphics[width=\textwidth]{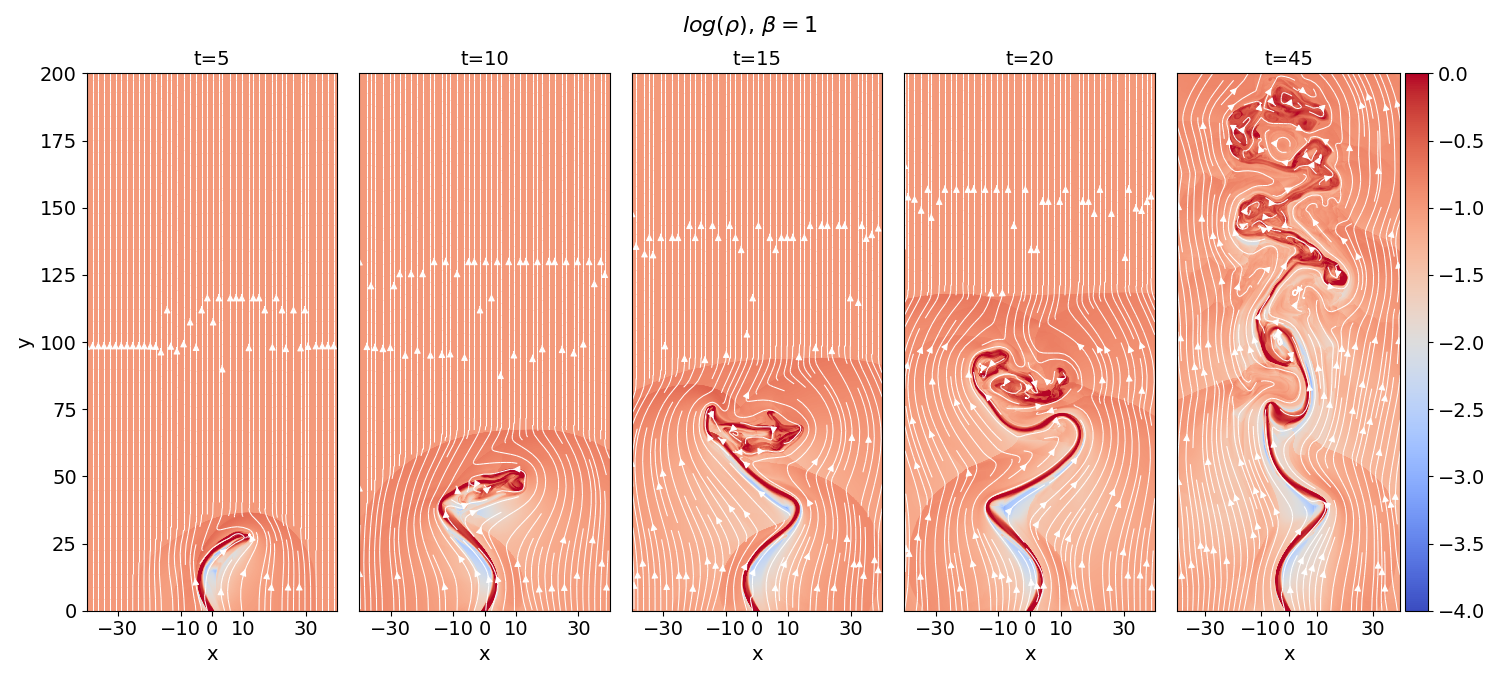}
\caption{Same as Figure \ref{fig:beta100_flines_rho} but $\beta=1$, simulation {\em pj30degB1}.
A larger grid is shown as the jet reaches farther distances.}
\label{fig:beta1_flines_rho}
\end{figure*}

\begin{figure*}
\centering
\includegraphics[width=\textwidth]{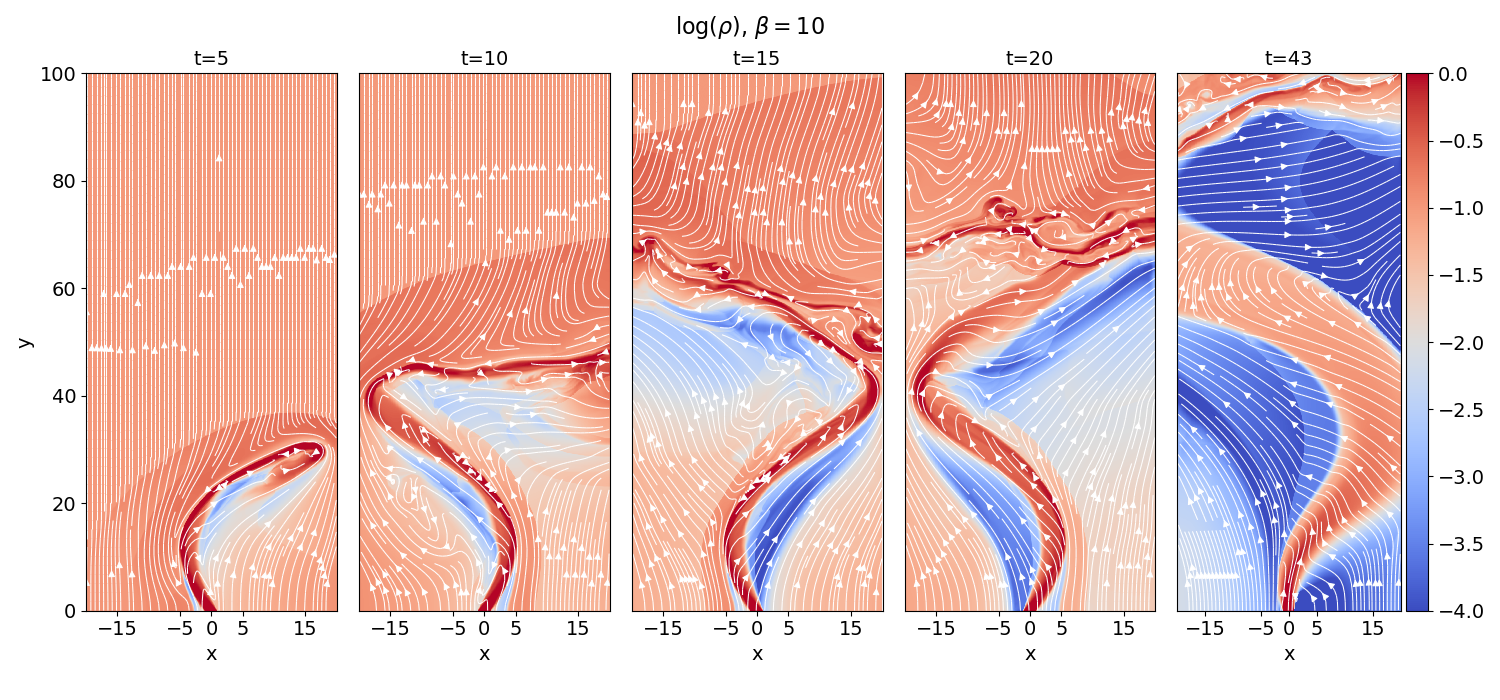}
\caption{Same as Figure \ref{fig:beta100_flines_rho} but $\beta=10$, simulation {\em pj30degB10}. }
\label{fig:beta10_flines_rho}
\end{figure*}

This can as well be seen in Figure~\ref{fig:beta1_flines_rho} and 
Figure~\ref{fig:beta10_flines_rho}, now applying a stronger magnetic field.
In particular for the simulation with $\beta =1$ ({\em pj30degB1}),
the evolution of the jet follows a rather straight trajectory, despite the same precession angle of the nozzle.
While the jet propagation follows initially (in time and space) a rather curved trajectory that is 
triggered by the injection nozzle, on longer spatial scales the jet propagation will straighten,
and will finally reach substantial longer distances compared to lower field strength.

For very long simulation times ($t>200$, not shown), we see an interesting change in jet geometry such
that the jet amplitude widens again as the ambient gas becomes diluted and with it the ambient
pressure that confines the jet.
For an intermediate field strength applying $\beta=10$ ({\em pj30degB1}) the
field is not sufficiently strong in order to collimate the wiggling jet flow such that if
follows a narrow channel and can reach large distance. 

\subsection{The force balance across the jet}
In order to understand the motions in the system of jet and ambient medium, we can look at the force 
balance across the jet
distribution across the system.
As the pressures do not remain constant during the evolution, forces arise that affect the jet propagation
and the distribution of the ambient gas.

\begin{figure}
\centering
\includegraphics[width=0.45\textwidth]{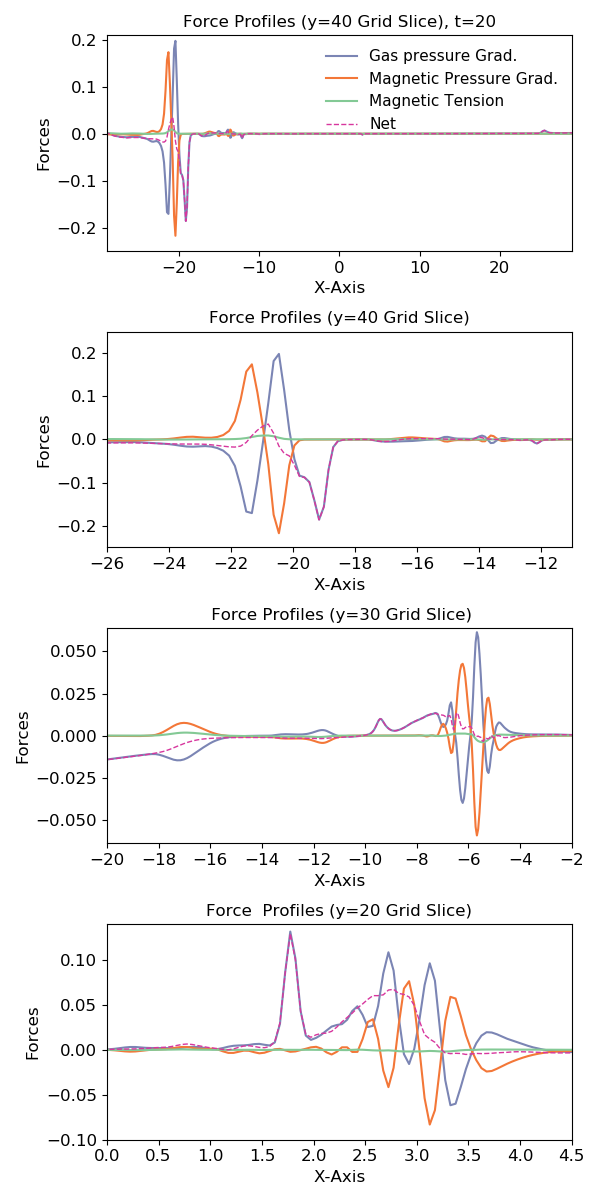}
\caption{The force profiles with different scales and different y-grid slices for $t = 20$ for simulation 
{\em pj30degB100}. 
The top panel shows forces along the $x$-direction at $y = 40$ for the whole x domain, $30 \leq x \leq -30$. 
In the panels below the force profile are shown, zoomed-in to the area around the jet for $y = 40, 30, 20$, 
respectively.
Shown are the gas pressure gradient (blue), the magnetic pressure gradient (orange), the magnetic tension 
(green), and the net force
(dashed pink).}
\label{fig:force-balance}
\end{figure}

In Figure~\ref{fig:force-balance} we show the profiles for the different force contributions.
Here, we consider simulation {\em pj30degB100} with a high plasma-beta $\beta = 100$, 
at time $t=20$ and for different layers across 
the jet, $y=20,30,40$ (see also Figure~\ref{fig:beta100_flines_rho} for a comparison).
We display the pressure gradient $\partial_x P$, the magnetic pressure gradient $\partial_x (B^2/2)$, 
and the magnetic tension $\vec{B} \cdot \nabla) \vec{B}$ which in x-direction is
$B_{x} \partial_{x} B_{x} + B_{y} \partial_{y} B_{x}$.

The profile drawn across the jet at altitude $y= 40$ (top panel) shows the localized jet disturbance that is 
embedded in a force-free ambient magnetized gas that is basically unchanged from the initial state.
In higher resolution of the jet cross-section, again at $y= 40$ (2nd top panel), we see that the force 
components in the jet almost balance, indicated by the dashed pink line.
However, just outside the jet (right side) the negative pressure rises again and remains unbalanced, giving 
rise to a force to the {'}left{'} direction, thus pushing the jet to the left. 

Going down along the jet to $y=30$ we see similar force profiles with the typical sinusoidal structure.
Compared to the $y=40$ slice, here an unbalanced (and positive) force on the left side of the jet appears, 
resulting again from gas pressure and pushing the jet to the right.

This is similar to a slice at $y=20$, however, here the gas pressure force is relatively larger.
We suggest that these high pressure forces arise since we are close to the jet injection nozzle where the 
jet is injected into the domain, in different directions over time, causing an over pressure.
On the contrary, in the upper areas, an under-dense area is established behind the jet when the jet is
moving left or right. 
The negative pressure force decelerates the lateral jet motion.

In all example slices, the area of gas pressure force is similar to the jet area.
In general, the magnetic tension force is relatively small compared to the magnetic pressure force.
While gas pressure gradients close to the jet affect the lateral jet motion, 
magnetic pressure force and gas pressure force almost balance within the jet (while changing sign).

\begin{figure}
\centering
\includegraphics[width=0.8\linewidth]{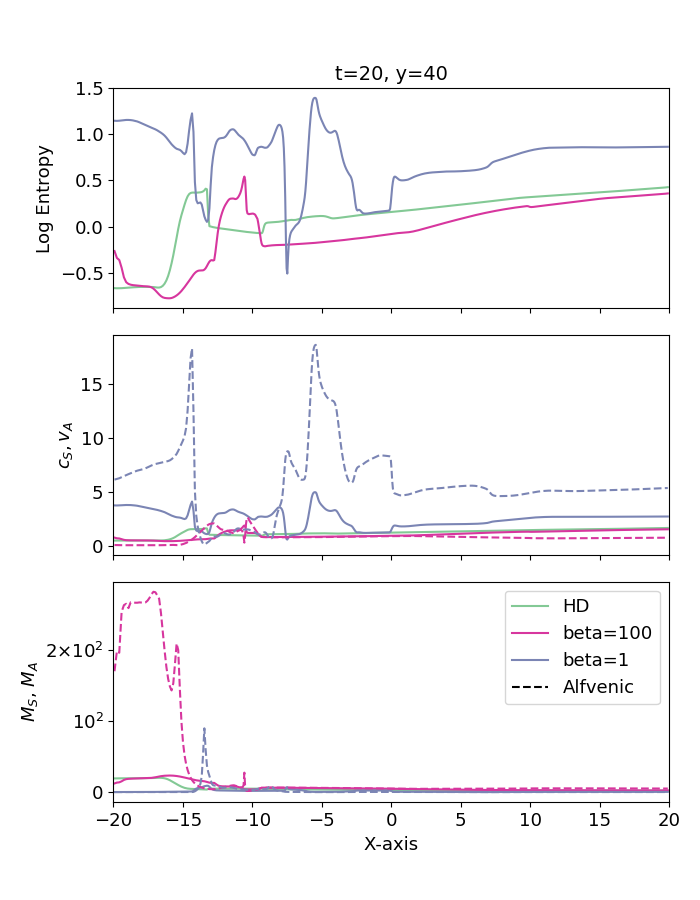}
\caption{(Magneto-)hydrodynamic key quantities shown as profiles across the jet for simulations 
{\em pj30deg} (hydro, green), 
{\em pj30degB100} (MHD, $\beta=100$, pink), and
{\em pj30degB1} (MHD, $\beta = 1$, blue), respectively.
From top to bottom, we show the profiles of 
the measure of entropy $K \equiv P/\rho^{\gamma}$, 
sound speed $c_{\rm S}$, 
Alfv\'en speed  $V_{\rm A}$ (dashed), 
Sonic Mach number $M_{\rm S}$, and 
Alfv\'en Mach number $M_{\rm A}$ (dashed). 
}
\label{fig:mach_alf_ent}
\end{figure}

Figure~\ref{fig:mach_alf_ent} shows profiles of characteristic MHD variables from the jet interaction.
In particular, it demonstrates how different a strongly magnetized jet behaves.
While the entropy measure ($K=P/\rho^\gamma$) is similar for the hydro case and the case of low magnetization, 
the highly magnetized jet ($\beta = 1$) is heated much higher temperatures.
Note that the hydrodynamics of the jet injection is the same for all models, only a force-free vertical field is
added for the magnetic case.

This additional magnetic energy serves as another energy source for jet heating, as
the $\beta=1$ jet carries an additional magnetic energy reservoir that is 100 times larger
than for the $\beta=100$ jet.
On the other hand, the Alfv\'en Mach number of the strongly magnetized jet is much lower compared to the weakly
magnetized jet (while the Alfv\'en speed is much higher).
This, of course, would have consequences for the stability of jet propagation, which we, in our 2D approach, cannot 
determine.

\subsection{Notes on a magnetized orbiting nozzle}
As mentioned above, we have concentrated our discussion on models considering a {\em precessing} jet nozzle.
The main effects of the magnetic field we just described are basically the same
for an orbiting nozzle, that is the damping of the jet curvature (see Fig.~\ref{fig:orbiting_nozzle_B} in the 
Appendix).

We close this section, noting that the magnetic field has quite an impact on how the binarity effects influence 
the jet motion.
The (longitudinal) magnetic field damps the lateral jet amplitudes that are enforced by either precession of orbital
motion of the jet sources.
Concerning jet thermodynamics, the magnetic field seems to serve as an additional energy reservoir that
leads to higher temperatures of the has that is shocked at the jet head or at the jet-ambient gas boundary layer.

\section{Conclusions}
\label{sec:sum}
We have applied 2D (M)HD simulations in order to investigate how a precessing or orbiting jet nozzle
affects the propagation of a high speed jet.
We have performed a parameter study of systems with
(i) different precession angle,
(ii) different orbital period, 
(iii) different orbital separation, and
(iv) different magnetic field strength.

While a modeling of exact astrophysical parameters is too CP-expensive for our parameter study,
our results can be applied for a number of astrophysical systems, such as close or wide binaries,
young stars, symbiotic stars, or compact binaries.
Our results are as follows.

(i) Both precession jet nozzles and orbiting jet nozzles inject jets that propagate on a curved trajectory.
The amplitude and the wavelength of the curved pattern depend on these initial parameters, such as that a
wider orbit or a wider opening angle leads to a larger amplitude.

(ii) While jets injected along orbits of large radius produce jets with large amplitudes, the jet motion that 
is injected by a precessing nozzle is limited by the opening angle of the precession cone. 
We find a {\em limiting opening angle} for the precession cone of about $20\degr$, above with the longitudinal
jet motion is too perturbed by precession in order to reach large distances.
These jets turn around following a strong side-wise motion and terminate soon in a broad shock region.

(iii) An essential difference between jets from precessing nozzles and orbiting nozzles is that on intermediate
length scales, the former show a (bipolar) S-shaped structure, while the latter show a C-shaped structure.
Note that due to our 2D approach, this has to be compared to a more realistic 3D structure, projected on
the plane of the sky.
However, the main features of the S-shape or C-shape will remain, and may be applied to observed sources.

(iv) While on intermediate distances a curved jet structure is revealed, on longer distances, thus also on
longer time scales, the persistent jet motion bores a funnel into the ambient medium, wide enough that the
jet wave pattern is enclosed.
On these length scales, the wave pattern dissolves into a broad stream filling smoothly the conically shaped
outflow funnel.
The outflow funnel is slightly oscillating, triggered by the time-dependent jet injection geometry.

(v) The consideration of a magnetic field in the ambient medium and the jet (a jet aligned poloidal field)
shows the stabilizing effect of magnetic forces (for jets injected by precessing and orbiting nozzles).
While simulations with a plasma-$\beta$ of 100 do not differ much from the hydrodynamic case, simulations
with $\beta = 10$ or even $\beta =1$ produce jets that propagate much further, since the lateral amplitude
triggered by the jet injected is damped on intermediate and long distances.
The jets remain confined to small jet radii, and the jet lateral amplitude is damped by a factor 2.

(vi) Investigating the force balance across the jet shows the gas pressure and Lorentz force are in good
equilibrium in the ambient medium and in the jet - except the region just outside the jet at the concave
side. 
This net force is finally balanced by the dynamical pressure of the lateral motion.

We finally note that our study is limited by the 2D approach. Orbital and precessing jet nozzles do not only
move left-right as in our 2D study, but orbit {\em around}. 
The main 3D effect here would be a helical wiggling of the jet stream. 
However, we firmly believe that their projected geometry and dynamics can be very well approximated
by our 2D approach.
While studies in the literature have partly addressed these questions, a full, long-term 3D MHD study is
still missing, one of the reasons being the huge CPU resources needed.
Jet launching simulations in 3D MHD considering binary sources have only recently been published, restricted
in size to the Roche lobe and in time to 1-2 orbital periods.

\acknowledgements
Melis Yard{\i}mc{\i} acknowledges financial support by the Scientific and Technological Research Council 
of Turkey (T{\"U}B\.{I}TAK 2214-A International Research Fellowship Program for PhD Students).
Our simulations were performed on a 72-core astro-node of the Max Planck Institute for Astronomy. 
M.Y. is grateful to the Max Planck Institute for Astronomy for the kind hospitality during her visit.
M.Y. and C.F. thank Giancarlo Mattia for sharing his routine for integrating field lines.
We acknowledge a detailed report with many helpful and interesting suggestions by an unknown referee. 

\appendix
\counterwithin{figure}{section}

\section{Comparison with literature simulations}
\label{appendix:comparison}
A number of previous studies have investigated the evolution and propagation of jets that are ejected from orbiting 
or precession sources (see discussion in Sect.\ref{sec:intro}).
In this paper we claim that our 2D Cartesian setup allows to investigate a series of parameter runs which could not 
be done by previous simulations, in particular for the 3D simulations.
It is therefore helpful to provide a table that compares the main simulation characteristics of these simulations 
with our present setup (see Table~\ref{tbl:comparison}).
One particular difference is that we also investigate the impact of the magnetic field.

In the following, we compare our results of the initial propagation with selected studies from the literature.
This is not straightforward as different parameter setups and numerical grids were applied 
(see Table~\ref{tbl:comparison} for a comparison).

For the precessing nozzle, we compare our simulations to the 3D hydro simulations of \citet{1996MNRAS.282.1114C}.
Both approaches show a wider opening angle of the jet outflow cone for a wider precession cone (see their Fig.~1 
and our Fig.~\ref{fig:t70_dif_prec_param}).
Also, when the precession time scale is short, more wiggles are observed (same figures) - interestingly, for a 
similar precession angle ($12\deg$ vs. $10\deg$) the jet length before termination is similar with about three 
wavelength.
Due to the higher resolution, our simulations show more sub-structure, such as shocked gas in and around the jet. 
These features may be very relevant for observations.
Further, we see a mild widening of the jet beam along its path, while the \citet{1996MNRAS.282.1114C}
simulations show a rather constant jet diameter.
Both simulations are run with jets in pressure equilibrium with the ambient medium.
On the other hand, the overall agreement on the main kinematic structures provides confidence for our 2D slab 
approach.

The 2D Cartesian simulations of \citet{1995MNRAS.275..557B} also show shock structures along the outer sides of 
the curved jet trajectory.
The simulations run for about three precession periods, showing three jet {"}waves{"} at this point in time.
However, we don't know how the system will evolve further in time, for example, whether the jet is able to 
propagate further.
The jet stream looks more stable, maybe because it is injected with a substantial over pressure (unlike in 
\citealt{1996MNRAS.282.1114C} or our approach).

We may further compare our results to the 3D simulations of \citet{2002ApJ...568..733M} who investigated orbiting 
and precessing nozzles in the hydrodynamic limit.
Here we find for both simulation setups only one wave pattern along the jet.
However, probably due to the high numerical costs, these simulations were evolved only till one orbital or 
precession period, respectively.
Also here, the jet streams stay rather narrow, despite their over-pressure with respect to the ambient gas, 
and the low jet resolution with $\simeq 3$ cells per jet radius.

\begin{table*}
\caption{Comparison of our simulation characteristics with a selection of literature simulations in the. 
We compare the size of the numerical grid and its geometry, 
the resolution applied for the jet, the simulation run time, and other specific features of the approach. 
}
\begin{center}
\begin{tabular}{lllll}
\hline    \noalign{\smallskip}
 Authors             & Approach     & \# grid cells                   & \# cells/jet radius,  & Remarks \\  
                     &              &                                 & run time              &  \\     
 \noalign{\smallskip}    \hline \hline    \noalign{\smallskip}
Biro et al.          & HD,  2D slab & $450 \times 450$                & 10                   & adaptive grid, incl.\\  
                     &              &                                 &                      & radiation/chemistry \\     
\noalign{\smallskip}    \hline \noalign{\smallskip} 
Cliffe et al.        & HD,  3D cart & $128 \times 128 \times 128$,    &  5 or 10,           &  \\ 
                     &              & $256 \times 256\times 256$ max  & 1.5 prec periods    &  \\
\noalign{\smallskip}    \hline  \noalign{\smallskip} 
Masciadri \& Raga    & HD, 3D cart  & $64 \times 64 \times 512$ max   & 3                   & adaptive grid, \\  
                     &              &                                 & 3 prec/orb periods  & orb. radius $=2.2 R_{\rm jet}$\\
\noalign{\smallskip}    \hline  \noalign{\smallskip} 
Monceau-Baroux et al.& HD, 3D cart  & $72 \times 72 \times 144$  (base) &                   & Adaptive mesh refinement \\ 
                     &              & $4608 \times 4608 \times 9216$ (eff) & 68 prec periods  & \\
\noalign{\smallskip}    \hline  \noalign{\smallskip} 
Sheiknezami et al.   & MHD, 3D cart & $200 \times 200 \times 200 $    & 100                  & launching, no ambient gas, \\ 
                     &              &                                 & 1 orbital period     & small physical grid \\
\noalign{\smallskip}    \hline  \noalign{\smallskip} 
Present study        & MHD, 2D slab &  $400 \times 2000 $             & 20                   & 200 cells / orb.~radius  \\  
                     &              &  $1200 \times 4000$ max         & 65 prec / 35 orb periods &  \\
\noalign{\smallskip}    \hline                 
\end{tabular}
\end{center}
\label{tbl:comparison}
\end{table*}

\section{Magnetized jets from an orbiting nozzle}
\begin{figure}
\centering
\includegraphics[width=0.95\textwidth]{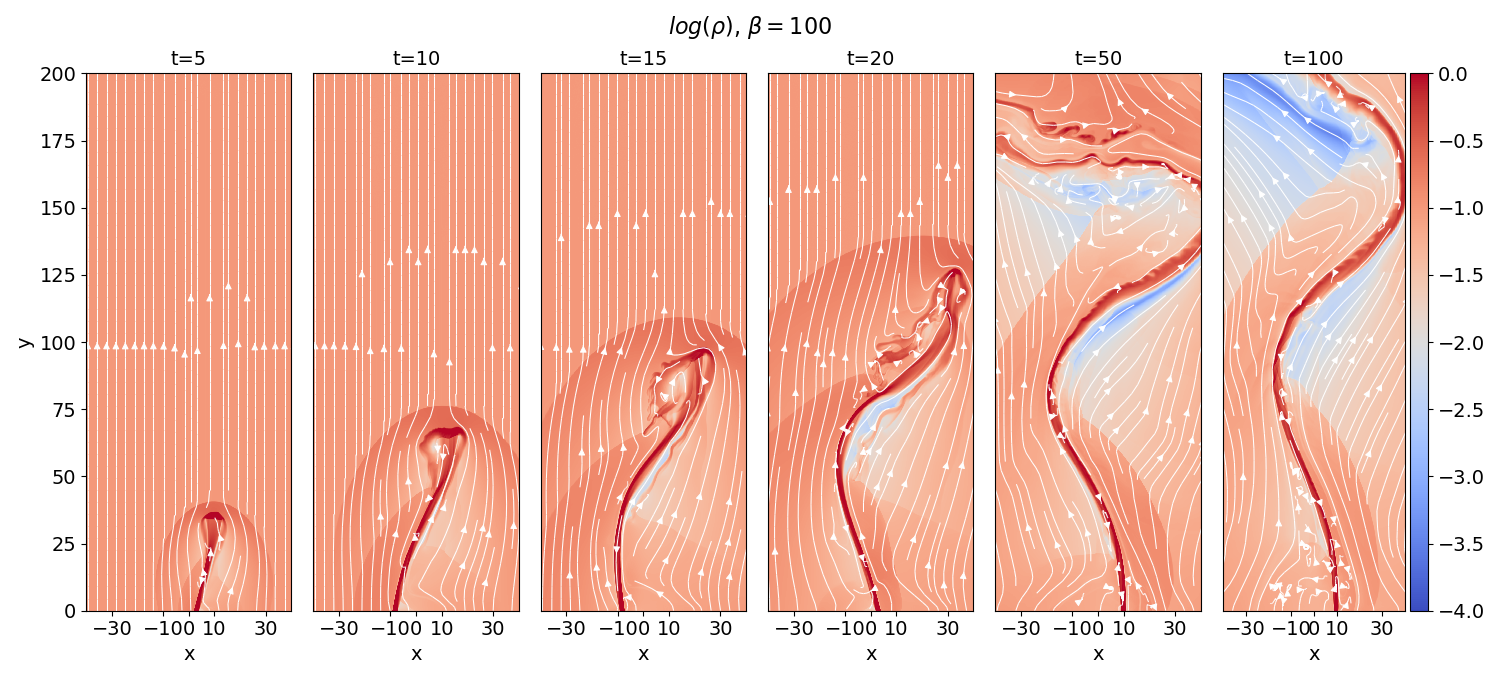}   
\includegraphics[width=0.45\textwidth]{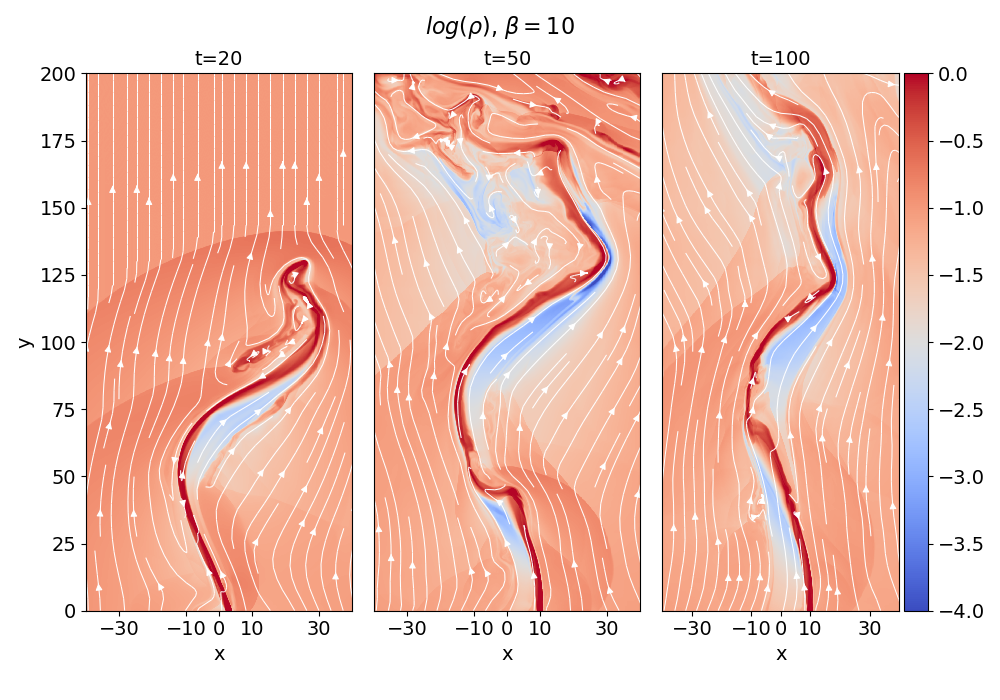}
\includegraphics[width=0.45\textwidth]{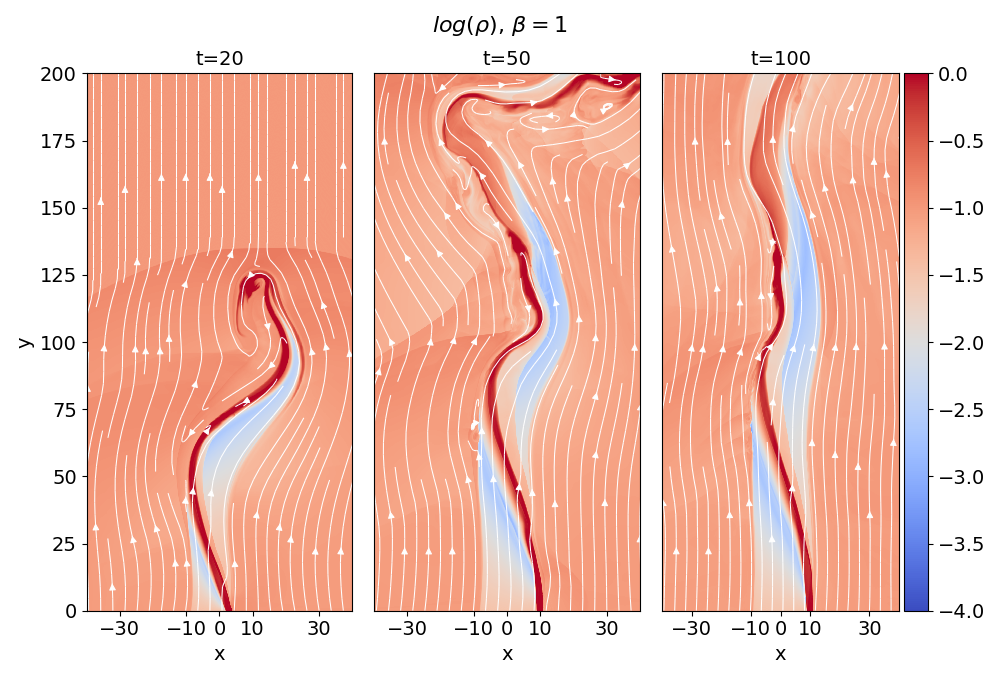}
\caption{Time evolution of the magnetized simulation for an orbiting jet nozzle 
{\em ojP25a10B100} with $\beta =100$ (top),
{\em ojP25a10B10} with $\beta=10$ (lower left), and
{\em ojP25a10B1} with $\beta=1$ (lower right),
all with orbital separation $a=10$ and an orbital period of $P=25$.
Shown is the density distribution (colors) and the magnetic field lines (white) 
for selected the time steps (left to right). 
} 
\label{fig:orbiting_nozzle_B}
\end{figure}

For comparison, we show here the evolution of the jet propagation for magnetized jets injected from an orbiting nozzle 
(see Fig.~\ref{fig:orbiting_nozzle_B}).
As it is clearly visible, the jet flow remains more stable in general.
As a consequence, it can reach farther distances in time.
The overall effect of the magnetic field on the curved jet structure is similar as for the precessing nozzle.

We can see its impact more quantitatively by comparing jets that are differently magnetized.
Comparing low magnetization ({\em ojP25a10B100}) with mild magnetization ({\em ojP25a10B10}) and with strong
magnetization ({\em ojP25a10B1}), the jet amplitude decreases from a $\Delta x \simeq 30$ to about $\Delta x \simeq 10$.
The small amplitude for the latter case is exactly corresponding to the orbital radius.
In comparison with the precessing nozzle of high magnetization (see above), the jet amplitude is even more damped.

While we do not expect a highly magnetized environment for protostellar jets and would thus expect relatively
large amplitudes for orbiting jets, the relativistic jets are launched in a highly magnetized environment
for which we expect a more straight propagation.

\bibliography{main}{}
\bibliographystyle{aasjournal}
\end{document}